\documentclass[useAMS,usenatbib]{mn2e}

\usepackage{graphicx} 
\usepackage{dcolumn}  
\usepackage{bm}       
\usepackage{amssymb,amsmath}  
\usepackage{color}
\usepackage{float}
\usepackage{natbib}
\newcommand{\ahf}{\textsc{AHF}}

\newcommand{\subfind}{\textsc{SUBFIND}}

\newcommand{\voboz}{\textsc{VOBOZ}}

\newcommand{\adaptahop}{\textsc{AdaptaHOP}}

\newcommand{\hotthreed}{\textsc{HOT3D}}
\newcommand{\hotsixd}{\textsc{HOT6D}}

\newcommand{\rockstar}{\textsc{Rockstar}}
\newcommand{\hbt}{\textsc{HBT}}
\newcommand{\hsf}{\textsc{HSF}}
\newcommand{\stf}{\textsc{STF}}

\newcommand{\skid}{\textsc{SKID}}
\newcommand{\grasshopper}{\textsc{GRASSHOPPER}}

\def\Mpc{\, h^{-1} \, {\rm Mpc}}
\def\kpc{\, h^{-1} \, {\rm kpc}}
\def\Mo{\, h^{-1} \, {\rm M_{\odot}}}
\def\kms{{\rm km}~{\rm s}^{-1}}

\begin{document}

\title[Subhaloes gone Notts]{Subhaloes gone Notts: the clustering properties of subhaloes}

\author[Arnau Pujol, Enrique Gazta\~{n}aga, Carlo Giocoli et al. ]{\parbox{\textwidth}{Arnau Pujol\thanks{E-mail: pujol@ice.cat}$^{1}$, Enrique Gazta\~{n}aga$^{1}$, Carlo Giocoli$^{2,3,4}$, Alexander Knebe$^{5}$, \\ Frazer~R.~Pearce$^{6}$, Ramin~A.~Skibba$^{7}$, Yago Ascasibar$^{5}$, Peter Behroozi$^{8,9,10}$, \\
Pascal Elahi$^{6,11,12}$, Jiaxin Han$^{11,13}$, Hanni Lux$^{6,14}$, Stuart~I.~Muldrew$^{6}$, \\
Mark~Neyrinck$^{15}$, Julian~Onions$^{6}$, Doug~Potter$^{16}$, Dylan Tweed$^{17}$ }\vspace{0.4cm}\\
$^{1}$Institut de Ci\`{e}ncies de l'Espai (ICE, IEEC/CSIC), E-08193 Bellaterra (Barcelona), Spain\\
$^{2}$Dipartimento di Fisica e Astronomia, Universit\`a di Bologna, viale Berti Pichat 6/2, 40127, Bologna, Italy \\
$^{3}$INAF - Osservatorio Astronomico di Bologna, via Ranzani 1, 40127, Bologna, Italy \\
$^{4}$INFN - Sezione di Bologna, viale Berti Pichat 6/2, 40127, Bologna, Italy\\
$^{5}$Departamento de F\'isica Te\'{o}rica, M\'{o}dulo 15, Facultad de Ciencias,
  Universidad Aut\'{o}noma de Madrid, 28049 Madrid, Spain\\
$^{6}$School of Physics \& Astronomy, University of Nottingham, Nottingham, NG7 2RD, UK\\
$^{7}$Center for Astrophysics and Space Sciences, Department of Physics, University of California, 9500 Gilman Drive, San Diego, CA 92093, USA\\
$^{8}$Kavli Institute for Particle Astrophysics and Cosmology, Stanford, CA 94309, USA\\
$^{9}$Physics Department, Stanford University, Stanford, CA 94305, USA\\
$^{10}$SLAC National Accelerator Laboratory, Menlo Park, CA 94025, USA\\
$^{11}$Key Laboratory for Research in Galaxies and Cosmology, Shanghai Astronomical Observatory, Shanghai 200030, China\\
$^{12}$Sydney Institute for Astronomy, University of Sydney, Sydney, NSW 2006, Australia\\
$^{13}$Institute for Computational Cosmology, Department of Physics, Durham University, South Road, Durham DH1 3LE, UK\\
$^{14}$Department of Physics, University of Oxford, Denys Wilkinson Building, Keble Road, Oxford, OX1 3RH, UK\\
$^{15}$Department of Physics and Astronomy, Johns Hopkins University, 3701 San Martin Drive, Baltimore, MD 21218, USA\\
$^{16}$University of Zurich, Institute for Theoretical Physics, Winterthurerstrasse 190, CH-8057 Zurich, Switzerland\\
$^{17}$Racah Institute of Physics, The Hebrew University, Jerusalem 91904, Israel \\
}
\date{Accepted xxxx. Received xxx}

\pagerange{\pageref{firstpage}--\pageref{lastpage}} \pubyear{2012}

\maketitle

\label{firstpage}

\begin{abstract}

We present a study of the substructure finder dependence of subhalo clustering in the Aquarius Simulation. We run 11  different subhalo finders on the haloes of the Aquarius Simulation and we study their differences in the density profile, mass fraction and 2-point correlation function of subhaloes in haloes. We also study the mass and $v_{\rm max}$ dependence of subhalo clustering. As the Aquarius Simulation has been run at different resolutions, we study the convergence with higher resolutions. We find that the agreement between finders is at around the $10\%$ level inside $R_{200}$ and at  intermediate resolutions when a mass threshold is applied, and better than $5\%$ when $v_{\rm max}$ is restricted instead of mass. However, some discrepancies appear in the highest resolution, underlined by an observed resolution dependence of subhalo clustering. This dependence is stronger for the smallest subhaloes, which are more clustered in the highest resolution, due to the detection of subhaloes within subhaloes (the sub-subhalo term). This effect modifies the mass dependence of clustering in the highest resolutions. We discuss implications of our results for models of subhalo clustering and their relation with galaxy clustering.

\end{abstract}

\begin{keywords}
methods: N-body simulations - methods: numerical - galaxies: haloes - cosmology: theory - 
\end{keywords}

\maketitle

\section{Introduction}

Large-scale structure in the Universe arises through the 
gravitational clustering of matter.  In the $\Lambda$CDM paradigm of 
hierarchical structure formation, gravitational evolution causes dark matter 
 to cluster around peaks in the initial density field and collapse later 
into virialized objects (haloes). These systems provide the potential well in which 
galaxies subsequently form \citep{White1978}.  It is therefore expected that 
the properties of a galaxy are correlated with the properties of its host halo.  
Small haloes merge to form larger and more massive haloes, which tend to be 
located in dense environments and are expected to host groups of galaxies, such 
that halo substructures are associated with satellite galaxies. 

Over the past few decades, numerical simulations have increased in size and 
resolution, and analytic models have become more sophisticated, such that the 
abundances of haloes (e.g., \citealt{Sheth1999}; \citealt{Warren2006}; \citealt{Tinker2008}), 
their clustering properties (e.g., \citealt{Mo1996}; \citealt{Sheth2001}; \citealt{Tinker2010}), 
assembly histories (e.g., \citealt{Giocoli2012}; \citealt{Neistein2010}; \citealt{Wechsler2002}), 
density profiles (e.g., \citealt{Navarro1997}; \citealt{Moore1999}), 
concentration-mass relations (e.g., \citealt{Maccio2007}; \citealt{Neto2007}; \citealt{Duffy2008}), 
and other correlations between their properties (e.g., \citealt{AvilaReese2005}; \citealt{Skibba2011}; \citealt{Wong2012}) are now better understood. 

Three rather different types of dark matter halo models have been developed 
to describe the connections between haloes and galaxies, and to explain the 
spatial distribution and clustering of galaxies in the context of hierarchical 
structure formation.  One class of models has come to be known as the `halo 
occupation distribution' (HOD; \citealt{Jing1998}; \citealt{Benson2000}; \citealt{Seljak2000}; \citealt{Scoccimarro2001}; \citealt{Berlind2002}), and describes how 
`central' and `satellite' galaxies of a particular type are distributed in 
haloes as a function of mass.  Complementary to this is the `conditional 
luminosity function' (CLF; \citealt{Peacock2000}; \citealt{Yang2003}; \citealt{Cooray2006}; \citealt{vdBosch2007}), which are based on a description of the 
luminosity (or stellar mass) distribution of galaxies as a function of halo 
mass. 


With improved numerical simulations, the abundances and properties of halo 
\textit{substructures} are being analysed with increasing precision, such that 
they can be reliably associated with (satellite) galaxies in groups and clusters 
(e.g., \citealt{Hearin2013}; \citealt{Reddick2013}), and can be modeled 
analytically as well \citep{Sheth2003,Giocoli2010}.  These 
developments have given rise to `subhalo abundance matching' models, with the 
unfortunate acronym, SHAM (e.g., \citealt{Conroy2006}; \citealt{Behroozi2010}; 
\citealt{TrujilloGomez2011}).  SHAMs typically assume a monotonic relation 
(perhaps with some scatter) between a galaxy property (luminosity or stellar 
mass) and a subhalo property (mass or maximum circular velocity, $v_{\rm max}$). 
By matching abundances of galaxies and haloes, these yield a description of the 
ways in which galaxies occupy haloes.  

Nonetheless, many uncertainties remain, and there is a need to better understand 
the distribution and clustering properties of subhaloes, and to quantify their 
systematics and biases.  Implicit or explicit assumptions are made about when a 
halo becomes a subhalo, how subhaloes experience dynamical friction and tidal 
stripping, when a subhalo has become disrupted \citep{Giocoli2008,Vandenbosch2005}. 
Moreover, HOD and CLF models usually assume a NFW number density profile, which 
can be different than subhalo density profiles used in SHAMs (e.g., \citealt{Zentner2005}; \citealt{Wu2013}).  In addition, there are difficulties and 
disagreements for subhalo-finding algorithms, about how to identify low-mass 
(sub)haloes, and how to treat mass stripping, `ejected' subhaloes, and 
dynamically unrelaxed structures.  For these reasons, it is crucial to compare 
and analyse the properties and spatial distribution of subhaloes for different 
subhalo finders and 
resolutions.  

Several comparison projects have been undertaken in the last few years 
\citep{Knebe2011,Onions2012,Elahi2013,Knebe2013,Onions2013,Srisawat2013}, 
with the purpose of studying the differences between various halo and 
subhalo-finding algorithms.  These studies have found that different methods 
can yield significantly different properties and statistics of dark matter 
structures.  For example, \citealt{Knebe2011} analysed the discrepancies and 
uncertainties in the measured halo masses and $v_{\rm max}$, and found some 
disagreement in the properties of low-mass haloes. 

\cite{Onions2012} focused on the subhalo-finding algorithms and studied the 
cumulative mass function and $v_{\rm max}$ function of different subhalo 
finders (Fig.\ $3$ and $6$ from \citealt{Onions2012}).  A simple comparison of these 
two functions show that the scatter of the cumulative $v_{\rm max}$ of the 
different subhalo finders is smaller than that of the cumulative mass function, 
implying that the subhalo finders obtain better agreement on the measurements 
of $v_{\rm max}$ than mass.  This is due to the fact that mass is strongly 
affected by the definition of the edge and shape of subhaloes, while 
$v_{\rm max}$ is constrained in the inner parts of the subhaloes 
\citep{Tormen2004,Giocoli2010,Muldrew2011}.  \citet{Elahi2013} studied the 
detection of streams in some of the subhalo finders, \citet{Onions2012} focused 
on spins and a summary of these comparisons is reviewed in \citealt{Knebe2013}. Recently \citet{Srisawat2013} studied a comparison of different merger tree algorithms.

In this paper, we study the subhalo finders' agreements and disagreements in 
subhalo clustering statistics, and the implications of these results on 
models.  We use the haloes from the Aquarius Simulation \citep{Springel2008} 
and $11$ different subhalo finders from the literature to study how the density 
profile and the 2-point autocorrelation function (2PCF) of subhaloes are 
affected by the finder algorithm.  We analyse the mass and $v_{\rm max}$ 
dependence of these measurements as well.

The Aquarius simulations have been run using different levels of resolution, 
which allows us to study the resolution dependence of these measurements.  
As the lowest resolutions of this simulation are close to 
the actual resolutions of the large-scale simulations such as the Millennium 
Simulation \citep{Springel2005}, the highest resolutions of the Aquarius haloes 
yield information about the effects and changes that would result by improving 
the resolution of these cosmological simulations.  These effects have 
implications on the subhalo clustering and can therefore affect constraints on 
galaxy formation models and halo models of galaxy clustering, including SHAMs. 

Our paper is organized as follows.  We describe the Aquarius Simulation and 
subhalo-finding algorithms in Section~2.  In Section~3, we describe our 
methodology, including the simulation post-processing and clustering 
measurements.  We present our results in Section~4: subhalo density profiles, 
mass fractions, and correlation functions.  We provide comparisons of different 
subhalo finders and resolutions, and analyse the dependence on subhalo mass. 
We also add an appendix to study the dependence on circular velocity.  
In Section~5, we provide an analytic halo-model description 
of the subhalo clustering signal.  Finally, we end with the conclusions and a 
discussion of our results.  


\section{Simulation and halo finders}\label{sec:simulation}

\subsection{Simulation}

For the study presented here we use the data from the Aquarius
simulation project \citep{Springel2008}, which consists of a set of five Milky
Way-like haloes (labelled A,B,C,D and E, respectively) each simulated at five different mass resolution levels (numbered as 1,2,3,4 and 5, in decreasing mass resolution). The cosmology
used for these zoom simulations is the same as that used for the Millennium Simulation
\citep{Springel2005}, i.e. a $\Lambda$CDM cosmology with parameters
$\Omega_m = 0.25$, $\Omega_\Lambda = 0.75$, $h = 0.73$, $n = 1$ and
$\sigma_8 = 0.9$. All simulations were performed in a box size of side length $100 \Mpc$ and the
number and mass of the particles inside those objects depends on the five levels of
resolution. In table~\ref{tab:haloes} we summarize the most
important characteristics of the particular haloes from the Aquarius suite used for our study; for more details
we refer the reader to \citet{Springel2008}.

\begin{table}
\begin{center}
\begin{tabular}{*{5}{c}}
\hline
halo & $m_p$ & $M_{200c} $ & $R_{200c} $ & $c$\\
 & $[\Mo]$ & $[10^{12}\ \Mo]$ & $[kpc]$ & \\
\hline
Aq-A-1 & $1.712 \times 10^3$  & $1.839$ & $245.76$ & $16.11$\\
Aq-A-2 & $1.370 \times 10^4$ & $1.842$ & $245.88$ & $16.19$\\
Aq-A-3 & $4.911 \times 10^4$ & $1.836$& $245.64$ & $16.35$\\
Aq-A-4 & $3.929 \times 10^5$ & $1.838$ & $245.70$ & $16.21$\\
Aq-A-5 & $3.143 \times 10^6$ & $1.853$ & $246.37$ & $16.04$\\
Aq-B-4 & $2.242 \times 10^5$ & $0.835$ & $188.85$ & $9.02$\\
Aq-C-4 & $3.213 \times 10^5$ & $1.793$ & $243.68$ & $14.84$\\
Aq-D-4 & $2.677 \times 10^5$ & $1.791$ & $243.60$ & $9.18$\\
Aq-E-4 & $2.604 \times 10^5$ & $1.208$ & $213.63$ & $8.52$\\
\hline
\end{tabular}
\caption{Selection of properties of those haloes from the Aquarius Project
suite that have been used for the present study. The number in the names refers to the level of resolution of
the simulation (decreasing resolution with increasing number).
$m_p$ is the mass of the high-resolution particles in the respective simulation,
$M_{200c}$ is the mass of the halo enclosed
within its radius $R_{200c}$, which in turn is the radius where the  mass overdensity
is $200$ times the critical density of the Universe. Finally, $c$ shows the
concentration parameter obtained from a fit to Navarro-Frenk-White \citep{Navarro1996} profile.}
\label{tab:haloes}
\end{center}
\end{table}

\subsection{Subhalo Finders}
Several substructure finders have been run on each of the haloes
listed in table~\ref{tab:haloes}.  These produce different subhalo
catalogues, with the differences obviously due to the different methods that the
subhalo finders use to find substructure within the dark matter
distribution of a halo. The same post-processing pipeline has been run to all the finders in order to make fair comparisons, as will be explained in \S \ref{sec:post_processing}. Our study aims at analysing the consequences of
these differences on the radial distribution and 2-point correlation
function of subhaloes. In this sub-section we provide 
a brief summary of the mode-of-operation of each of these codes. For more details and actually additional comparisons we refer the reader to various other papers dealing with the Aquarius data set and emerging as a result of our "Subhalo Finder Comparison Project", respectively \citep[e.g.][]{Onions2012,Onions2013,Knebe2013}.

\subsubsection{\adaptahop}

\adaptahop \citep{Tweed2009} starts by finding smoothed local density
peaks. Subhaloes are then found according to a hierarchical tree
obtained from the saddle points formed by increasing a density
threshold. This finder is purely topological: it does not use any
unbinding process for the particles associated with each subhalo.

\subsubsection{\ahf}

The halo finder \ahf\footnote{\ahf\ is freely
  available from
  \texttt{http://www.popia.ft.uam.es/AHF}}
  \citep[\textsc{amiga} Halo Finder,][]{Gill2004,Knollmann2009}
 is a spherical overdensity finder that simultaneously
identifies isolated haloes and sub-haloes. The initial particle lists are obtained
by a rather elaborate scheme: for each subhalo the distance to its
nearest more massive (sub-)halo is calculated and all particles within
a sphere of radius half this distance are considered prospective
subhalo constituents. This list is then pruned by an iterative
unbinding procedure using the (fixed) subhalo centre as given by the
local density peak determined from an adaptive mesh refinement
hierarchy. 

\subsubsection{\hbt}

Hierarchical Bound Tracking \citep[\hbt, ][]{Han2011} obtains the
subhaloes of Friends of Friends (FOF) groups by studying their merger
trees and identifying the remnants of smaller FOF groups that have
merged or been accreted. \hbt\ is a tracking finder, in that it
requires the previous history of any present structures to be known.

\subsubsection{\hotthreed\ \& \hotsixd}

\hotthreed\ and \hotsixd\ compute the Hierarchical Overdensity Tree (HOT) in
an arbitrary multidimensional space. It is analogous to the minimal
spanning tree (MST) for Euclidean spaces, but using the field obtained
from the FiEstAS (Field Estimator for Arbitrary Spaces) algorithm
\citep{Ascasibar2005,Ascasibar2010}. \hotthreed\ identifies density maxima
in configuration space, while \hotsixd\ identifies maxima in full six
dimensional phase-space.

\subsubsection{\hsf}

The Hierarchical Structure Finder \citep[\hsf, ][]{Maciejewski2009}
identifies all the particles to a given phase-space density maxima
above a certain density threshold by following the gradient of the
phase-space density field. After this first association, all the
particles that are gravitationally unbound to the corresponding maxima
are removed from the final substructure object.

\subsubsection{\grasshopper}

\grasshopper\ (Stadel, in prep.) is a reworking of the \skid\ group
finder \citep{Stadel2001}. It finds density peaks in the field and all
the particles bound to them. Particles are slowly slid along the local density gradient until they pool at a maximum, each pool corresponding to each initial group. Each pool is then unbound by iteratively evaluating the binding energy of every particle in their original positions and then removing the most non-bound particle until only bound particles remain.

\subsubsection{\rockstar}

\rockstar\ \citep[Robust Overdensity Calculation using K-Space Topologically
Adaptive Refinement,][]{Behroozi2013} is a recursive FOF
algorithm. The first selection of particle groups comes from running a
FOF with linking length $b = 0.28$. For each main FOF group, \rockstar\ builds a
hierarchy of FOF subgroups in phase-space by progressively and
adaptively reducing the linking length, so that a tunable fraction
(70~per cent, for this analysis) of particles are captured at each subgroup
as compared to the immediate parent group. And eventually only gravitationally
bound particles are kept.

\subsubsection{\stf}

The STructure Finder \citep[\stf\ a.k.a. \textsc{VELOCIraptor, }][]{Elahi2011}
identifies objects by utilizing the fact that dynamically distinct substructures in a halo will have a {\em local} velocity distribution that differs significantly from the mean, {\em i.e.} smooth background of the halo. Dynamically distinct particles are linked using a FOF-like approach
and an unbinding procedure is applied. This finder allows the
detection not only of virialized subhaloes, but also tidal streams
that can come from disrupted subhaloes.

\subsubsection{\subfind}

\subfind\ \citep{Springel2001} starts with a standard FOF analysis. In each of
FOF groups the highest density peaks are found, and the saddle points are
located by decreasing the density threshold of these peaks. The subhalo
candidates are obtained from these saddle points, and any
gravitationally unbound particles are removed from these candidates.

\subsubsection{\voboz}

\voboz\ \citep[VOronoi BOund Zones, ][]{Neyrinck2005} is based on a Voronoi tessellation, from
where the density peaks are found. Each particle is associated with a
peak that lies up the steepest density gradient from the
particle. A statistical significance is measured for
each (sub)halo, based on the probability that Poisson noise would
produce it. Finally, gravitationally unbound particles are removed.

\section{Methodology}

\subsection{Post-processing}\label{sec:post_processing}

In  order  to   make  fair  comparisons  of  the   finders,  the  same
post-processing pipeline  has been  applied to  all of them. Each
finder provider was asked to run  their algorithm on the Aquarius haloes
and return a list of the  identified subhaloes with the particles that
belong to each of them.  From these particle lists the same
analysis has  been applied to  obtain the different properties  of the
subhaloes.   As  the  finders   present  different  methodologies  for
post-processing  the  halo  particles  (e.\  g.\  they  use  different
definitions of subhalo centre, mass or thresholds),
these differences could confuse the  finder comparison, since we would
not  be  able  to  distinguish   which  differences  are  due  to  the
substructure finder algorithm and which  ones are due to the different
criteria used in the post-processing. For this reason, all subhalo finders only returned 
particle ID list of the subhaloes, and these ID lists have been uniquely post-processed 
by one code as described in \cite{Onions2012}.
 


As level $4$ is the highest resolution where all the finders have been
run, this  is the level  of resolution that we  will use in  our study
when we compare  all the finders. Only $3$ finders  (\ahf, \rockstar\ and
\subfind) have been  run in all the  levels, so we will  focus on these
finders when we study the resolution dependencies of the measurements.

\begin{table*}
\begin{center}
\begin{tabular}{*{7}{c}}
\hline
Finder & \multicolumn{2}{|c}{$M > 2 \times 10^7 \Mo$ } & \multicolumn{2}{|c}{$M > 10^8 \Mo$ } & \multicolumn{2}{|c}{$v_{\rm max} > 10\ \kms$}\\
& \multicolumn{1}{|c}{$r < 500\kpc$} & $r < R_{200}$ & \multicolumn{1}{|c}{$r < 500\kpc$} & $r < R_{200}$ & \multicolumn{1}{|c}{$r < 500\kpc$} & $r < R_{200}$\\
\hline 
ADAPTAHOP & $1744$ & $1329$ & $299$ & $213$ & $422$ & $322$\\
AHF & $1146$ & $624$ & $279$ & $155$ & $535$ & $339$\\
HBT & $1087$ & $588$ & $262$ & $143$ & $530$ & $334$\\
H3D & $1009$ & $583$ & $250$ & $147$ & $514$ & $337$\\
H6D & $941$ & $572$ & $250$ & $147$ & $496$ &$331$\\
HSF & $1064$ & $585$ & $260$ & $144$ & $518$ & $328$\\
GRASSHOPPER & $1070$ & $583$ & $258$ & $146$ & $538$ & $337$\\
ROCKSTAR & $1207$ & $629$ & $290$ & $157$ & $551$ & $350$\\
STF & $960$ & $563$ & $224$ & $134$ & $478$ & $309$\\
SUBFIND & $964$ & $549$ & $232$ & $133$ & $488$ & $315$\\
VOBOZ & $1191$ & $635$ & $245$ & $135$ & $514$ & $342$\\
\hline
\end{tabular}
\caption{Number of subhaloes found in the Aq-A-4 halo for each subhalo
  finder at  different mass  and $v_{\rm max}$  thresholds. They  are also
  compared to the same thresholds but restricted to $r < R_{200}$.}
\label{tab:n_sub}
\end{center}
\end{table*}

In Table  \ref{tab:n_sub} we  show the number  of subhaloes  found for
each finder in  the Aq-A-4 halo with different  thresholds.  The first two columns 
show the number of subhaloes more massive than $M > 2 \times 10^7\ \Mo$,
 where the first represents all subhaloes within $r<500 \kpc$ and the second all 
 subhaloes whose center lies within $R_{200}$.  The  following two columns show the same  but for a
mass threshold of $M > 10^8  \Mo$.  Finally, the last two columns show
the  number of  subhaloes with  $v_{\rm max} >  10\ \kms$,  at $r < 500 \kpc$ and
 $r < R_{200}$ respectively. As our post-processing pipeline is restricted to $r < 500 \kpc$,
  the first column of each threshold corresponds to all the subhaloes found 
  in the halo for this threshold. The amount of substructure outside $R_{200}$
 depends strongly of the algorithm. Although these overdensites are found, some finders 
 consider them as subhaloes while other define them as haloes. For this reason we will 
 restrict our analysis to the subhaloes inside $R_{200}$. From the table, we notice an excess of
\adaptahop\ subhaloes at low masses;    as    consistently    discussed    in
\citealt{Onions2012}. This is due to  the fact that \adaptahop\ does not
have  any  unbinding process  and  many  systems with  gravitationally
unbound particles  are considered subhaloes.  This  produces an excess
of small subhaloes in  the densest regions.

\subsection{Correlation Functions}\label{sec:corr_funcmethod}

We computed  the 2-Point Correlation  Function (2PCF) of  subhaloes in
the  Aq-A halo  for the  samples obtained  from the  different subhalo
finders and for  the different levels of resolution.  We also computed
the cross  correlation function (cross  CF) between the  subhaloes and
the centre  of the halo.  We compared  the behaviour of  the different
finders and the  dependence on the resolution level. As  the number of
subhaloes  found  in  each  finder   is  different  (see  table  2  in
\citealt{Onions2012}), we always used  thresholds in mass or $v_{\rm max}$
in order to compare similar samples.

The  2PCF can  be  obtained  by normalizing  the  number  of pairs  of
data-data as a function of distance ($DD(r)$) to the number of random-random pairs:
\begin{equation}
\xi_{ss}(r) =\frac{DD(r)}{RR(r)} - 1
\end{equation}
where $\xi_{ss}$  refers to the subhalo-subhalo  correlation function.
In this paper, we assume a uniform distribution with no border effects
instead of  using random  samples (thus the  above estimation  is then
equivalent  to the  commonly used  \citet{Landy1993} estimator).   The
number of  random-random pairs separated  a given distance $r$  can be
expressed as:
\begin{equation}
RR = \frac{1}{2} N_{s} n_{s} dV
\end{equation}
where $N_{s}$ and $n_{s}$ are the total number and mean densities of
the subhalo samples and $dV=\frac{4 \pi}{3} [( r + dr )^3 - r^3]$. The
expression of $\xi_{ss}(r)$ becomes:

\begin{equation}
\xi_{ss}(r) = \frac{2 DD}{N_{s} n_{s} dV} - 1,
\label{eq:xiss}
\end{equation}
Is very important  to mention that the  normalization of $\xi_{ss}(r)$
is  arbitrary  since it  depends  on  the  universal mean  density  of
subhaloes  $n_{s}$. The  mean  density  of the  halo  depends on  the
definition of its edge, and also the  mean density in a halo might not
be  representative of  the  mean  density of  the  Universe. As  large
simulations such as the  Millennium Simulation \citep{Springel2005}, with lower resolution,
cannot bring  information about the  abundance of small  subhaloes, we
define  the  mean  density  simply by  normalizing  the  abundance  of
subhaloes  to   a  volume   of  $1  (\rm Mpc/h)^3$,   so  $n_{s}   =  N_{s}
(\rm Mpc/h)^{-3}$.   Note  how  this   definition  makes  the  correlation
independent of  the total number of  subhaloes in each sample.  This is
important for  the relative comparison between  samples.  The $n_{s}$
normalization factor only  affects the overall amplitude  of the 2PCF,
and  also  can  distort  the  largest  scales  (at  the  edge  of  the
halo).  This  effect  is  important in  terms  of  global  statistical
implications,  but   it  does not   affect  comparisons   and  relative
values. As this 2PCF is only for subhaloes in one halo, this 2PCF must
be understood  as the contribution  that this  halo would have  to the
1-halo term in  a 2PCF of subhaloes of a  large and homogeneous volume
(with the corresponding amplitude).

The cross CF  (halo-subhalo CF) is estimated assuming  the same volume
and  densities, in  order to  be  compared with  the 2PCF.  As we  are
interested in the radial distances, we have no border effects and, therefore, 
we normalized the  number of halo centre-subhalo pairs  to the volume,
assuming a uniform distribution:
\begin{equation}
\xi_{hs}(r) = \frac{1}{n_{s}}\frac{N_{s}(r)}{dV} - 1
\label{eq:xi_hs}
\end{equation}
where $n_{s}$ is  the number density of subhaloes  and $N_{s}(r)$ is
the subhalo density profile: the number of subhaloes in a radial shell
of volume  $dV$ at distance $r$.   Again, this is the  contribution of
only one halo, where we assume  an arbitrary $n_{s}$ that has effects
in  the amplitude  but  not  in the  relative  comparisons.  Note  how
$\xi_{hs}(r)$ contains  the same information as  $N_{s}(r)$, but is just
normalized  as a  correlation function.  The interest  of showing  this
function  in   this  way   is  to  make   a  closer   comparison  with
$\xi_{ss}(r)$, which is the main object of our analysis.

\section{Results and comparisons}\label{sec:results}

\subsection{Measurement Comparison}\label{sec:meas_comp}

\begin{figure}
\begin{centering}
\includegraphics[width=84mm]{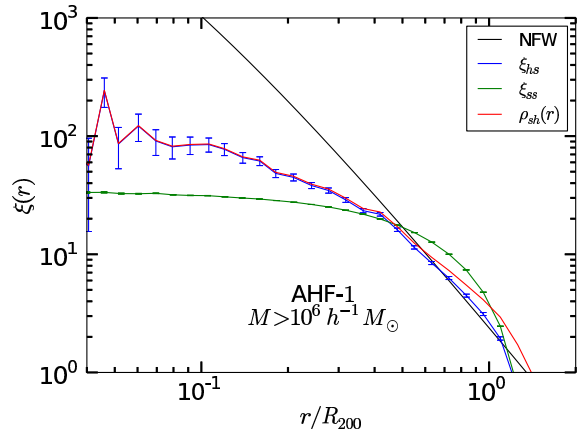}\caption[Comparison
  of  cross  CF vs  2PCF]{Comparison  between  the 2PCF  of  subhaloes
  (green),  the cross  CF  between  the centre  of  the  halo and  the
  subhaloes  (blue)  and  the  number density profile  (red)  in  Aq-A-1  for
  subhaloes with $M > 10^6 \Mo$ from AHF. The black line shows the NFW
  dark matter density  profile of the halo. The  normalization of this
  profile has been done to compare  it better with the other lines. To
  make a fair comparison between the number density profile and the cross CF,
  we have normalized the number density  profile by a factor $1/n_{s}$. Here
  $R_{200}     =      245.76     \kpc$. The errors of the correlation functions are from Poisson shot-noise.}     \label{fig:cross_vs_2pcf}
\par\end{centering}
\end{figure}

In Fig.\  \ref{fig:cross_vs_2pcf} we  show the comparison  between the
subhalo  2PCF (green) and the cross  CF between  subhaloes and  the halo
centre (blue); in  the figure we also show for  comparison the number density
profile  $\rho(r)$ (red)  of  the  subhaloes in  the  halo Aq-A.   The
subhaloes are those with  $M > 10^6 \Mo$ from the  AHF finder, but the
comparisons from other  finders are similar. 

First of  all, we observe
that the  number density profile and  the cross CF are  similar measurements,
since both are measuring the amount  of subhaloes as a function of the
radial distance to  the centre of the halo. To  make a fair comparison
we have normalized  the number density profile a factor  of $1/n_{s}$, since
this factor relates both magnitudes.  The only difference between them
is at the  edge of the halo,  where $\xi_{hs}$ is close  to unity, and
the factor $-1$  of equation \ref{eq:xi_hs} becomes  relevant.  

On the
other hand,  we can  see that $\xi_{ss}$  is flatter  than $\xi_{hs}$.
This is  expected since $\xi_{ss}$  is measuring the  distance between
pairs  of subhaloes  and  the  probability of  having  pairs at  large
distances must be larger that in  the case of $\xi_{hs}$, where one of
the pairs is  always in the centre. We show the shot-noise error bars of the 
correlation functions in order to show that the statistics of subhaloes is
 enough to trust their differences. Finally, the  black line shows the
NFW \citep{Navarro1996} fit of the dark  matter field of the halo (see
Table  $2$  of  \citealt{Springel2008}).  We  normalized  the  number density
profile to  make our  comparison more  visible. We  can see  that the
number density profile of subhaloes is very different than the NFW profile of
the dark matter,  at least for subhaloes in $r/R_{200}  < 0.4$ (around
$100  \kpc$ for  this particular  halo). One  of the  causes of  these
differences can be an effect of exclusion produced in the inner regions,
where the  core of the halo  dominates, the subhaloes  are easily
merged, and there are strong  tidal stripping effects that can disrupt
the subhaloes (and lose mass beyond  our sensitivity). Then, although  there is a
high density, it is difficult to find substructure. 

On the other hand,
understanding the difference between  subhalo and dark matter profiles
can be  a step forward  to galaxy  formation.  
In some halo models of galaxy clustering, such as HOD and CLF models (e.g., \citealt{Zheng2007}; \citealt{vdBosch2007}; \citealt{Zehavi2011}),  it is typically assumed 
that galaxies follow the dark matter distribution, with a NFW profile and 
a particular mass-concentration relation (e.g., \citealt{Maccio2008}). 
 Recent models also account for the fact that the distributions of subhaloes (e.g. \citealt{Gao2004}; \citealt{Klypin2011}) 
 and satellite galaxies (e.g. \citealt{Yang2005};
  \citealt{Wojtak2013}) appear to be less 
 concentrated than that of dark matter.  On the other hand, SHAM models directly 
 associate galaxies with identified subhaloes.  A potential difficulty is the treatment of 
 stripped, disrupted, or `orphan' satellites, in which the subhalo has been stripped and 
 is no longer resolved but the galaxy remains intact.  We discuss this further later in the paper.


\begin{figure}
\begin{centering}
\includegraphics[width=84mm]{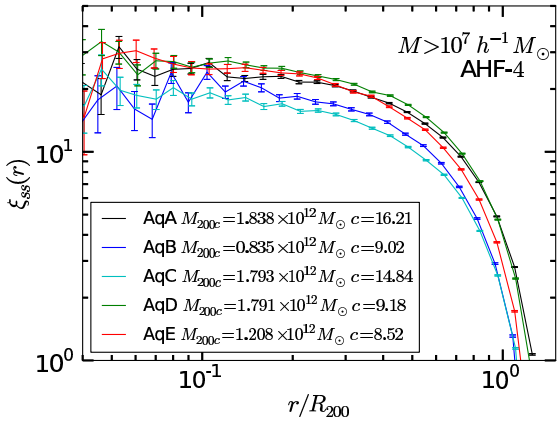}
\caption[Comparison of  2PCF for  different haloes]{Comparison  of the
  different haloes  of the  2PCF of  subhaloes for  the AHF  finder at
  level $4$ of resolution. The subhaloes have a mass threshold of $M >
  10^7\ \Mo$. Errors are from Poisson shot-noise.} \label{fig:cf_vs_halo} \par\end{centering}
\end{figure}

In Fig.\  \ref{fig:cf_vs_halo} we study the  difference between haloes
by measuring  the 2PCF  of subhaloes  for all  the Aquarius  haloes at
level  $4$  of  resolution.  We  show  the  measurement  for  the  AHF
subhaloes, although  the other  finders show  similar results.  A mass
threshold of  $M >  10^7\ \Mo$  has been  applied. We need to increase the mass 
threshold with respect to Fig.\ \ref{fig:cross_vs_2pcf} because the
 resolution in Fig.\ \ref{fig:cf_vs_halo} is lower. In order to compare the haloes, the same 
 normalization has been applied. We have assumed a mean number density of subhaloes 
 according to the number of subhaloes in the Aq-A halo, so $n_s = N_{s,\rm AqA}(\rm Mpc/h)^{-3}$. 
 As all the haloes belong to the same cosmology, they should have the same cosmological $n_s$. 
 Then, in Fig.\  \ref{fig:cf_vs_halo} we see the contribution of each halo to the 1-halo 
 term of the 2PCF of subhaloes in a large cosmological simulation, with an arbitrary normalization. 
 In this sense, haloes with more subhaloes will tend to contribute more strongly.  There  are clear
differences between  haloes that are not due to the lack of statistics, as 
shown from the shot-noise error bars (for clarity of the figures we will not
 show these errors in the rest of the figures of the study, but the magnitude of these 
 errors is the same in all the paper). Note that these differences are not caused by our arbitrary normalization of 
$n_{s}$ in Eq.\ref{eq:xiss}. The normalization has to be the same
for all haloes, as we are measuring the correlation with respect
some global (but unknown) universal mean density of subhaloes 
(above   $M >  10^7\ \Mo$). If we change the normalization
value all correlations will change by the same factor.

For the 2 cases where masses are very similar (C and D) the one
with larger concentration has lower amplitude. This is due to the fact that halo Aq-C 
has much less subhaloes than Aq-D, and then the contribution to the 1-halo term of the 2PCF is smaller. 
 This  is an indication that  the evolution or
other  properties  of  the  haloes  can  produce  differences  in  the
clustering although having the same mass. As the Aq-A halo is the only
one from where  the highest resolutions are available for  some of the
finders, we will focus on this halo in the rest of the study.

\subsection{Density profiles}\label{sec:dens_prof}

\begin{figure}
\begin{centering}
\includegraphics[width=84mm]{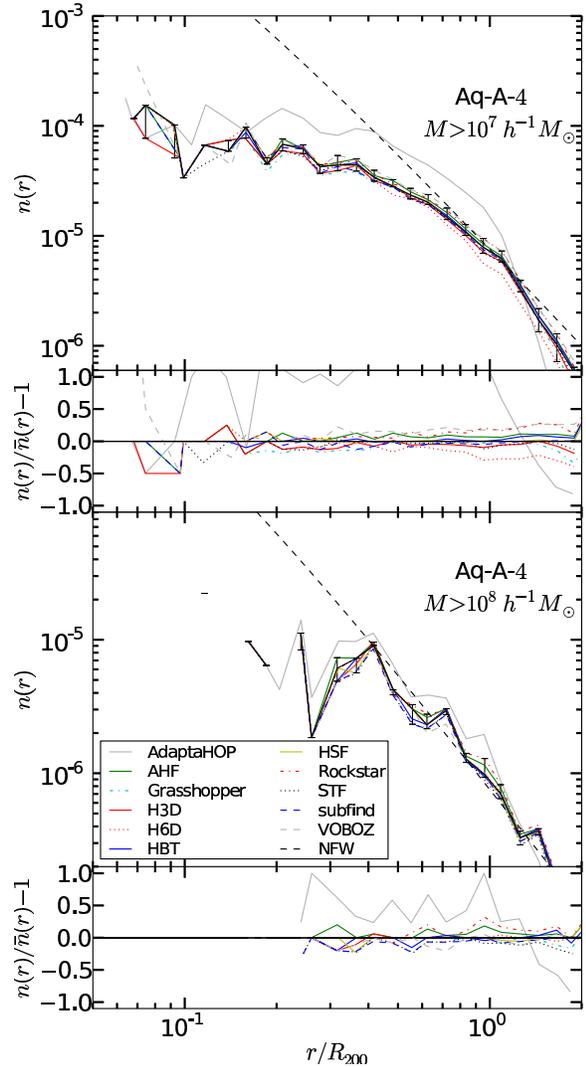}
\caption[Density profile of subhaloes in the Aq-A halo]{ Comparison of
the subhalo number density profile of the Aq-A halo at resolution
level $4$ for all the contributing finders. A mass threshold of $M >
10^7 \Mo$ (top) and $M > 10^8 \Mo$ (bottom) has been
applied. For Aq-A, $R_{200} = 245.70 \kpc$. The black dashed line represents de NFW fit 
of the dark matter density profile of the halo with an arbitrary normalization.}
\label{fig:dens_prof} \par\end{centering}
\end{figure}

In Fig.\ \ref{fig:dens_prof} we show the subhalo number density
profile of the Aq-A halo for each of the subhalo finders at the 4th
level of resolution. In the top panel, the number density profile is
restricted to subhaloes with $M > 10^7 \Mo$, while the bottom
panel shows the number density profile using a subhalo mass threshold of $M >
10^8 \Mo$. For most of the finders there is good agreement
however, especially when the smallest subhaloes are included, we can
see a large excess of subhaloes for \adaptahop\ with respect to
the rest of the finders. In
what follows this finder will often show differences with
respect to the others in the comparisons. This is largely
because, as discussed elsewhere \citep{Onions2012,Knebe2013} this
finder do not include a proper unbinding procedure in their subhalo
extraction process which can lead to an overdetection of subhaloes as
explained in \S \ref{sec:post_processing}. 

We can also see that the radial range of these
number density profiles depends upon the mass threshold. This is largely
because the more massive subhaloes are significantly rarer. Although
they appear to preferentially reside in the outer part of the halo
this is in fact not the case: the number density profile of the more
massive haloes is significantly steeper than the low mass subhaloes
and they are in fact, as expected due to dynamical friction, more
centrally concentrated than the low mass subhaloes.

As we can see from Fig.\ \ref{fig:dens_prof} the subhalo number
density profile is not only different from the underlying dark matter
density profile of the host halo but it also depends on the mass of
the subhaloes, being significantly steeper and so more centrally
concentrated for higher subhalo masses.  These effects could have
important consequences when trying to understand the distribution of
galaxies in haloes but would need to be investigated with a large
ensemble of host haloes spanning a broad range of mass and formation
history rather than the small number we have at our disposal here. A
study along these lines could be used to improve the models of subhalo
statistics from the halo model
\citep{Cooray2002,Sheth2003,Giocoli2010b}.

\begin{figure}
\begin{centering}
\includegraphics[width=84mm]{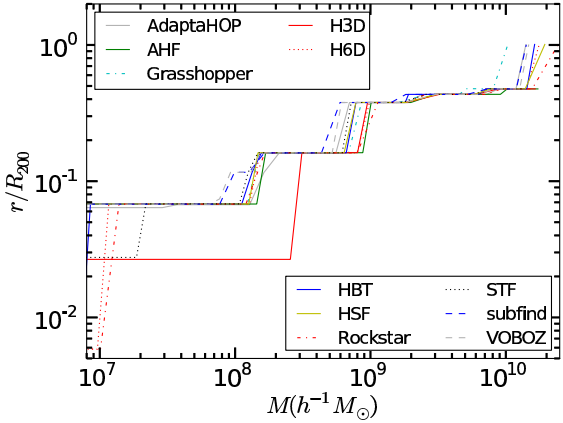}
\caption[Minimum radius of subhaloes as a function of mass]{Minimum
radial distance to the halo centre that a subhalo below the stated
mass first appears for all the different subhalo finders.}
\label{fig:r_min_vs_mass} \par\end{centering}
\end{figure}

In Fig.\ \ref{fig:r_min_vs_mass} we study how stable the different
finders are at recovering haloes of different masses as a function of
distance from the halo centre. We recover the minimum radial distance where subhaloes below the stated
mass first appear. A systematic offset in this figure would indicate a
finder that was struggling to find subhaloes close to the halo
centre. Due to their rarity more massive subhaloes are found farther
from the halo centre than small subhaloes. The agreement between
finders for the most massive subhaloes is remarkable but not exactly
surprising: such large objects far from the halo centre are easy to
spot.  The scatter between the finders is larger in the low mass
region where the presence or absence of a small object near the halo
centre can make a difference.

\begin{table*}
\begin{center}
\begin{tabular}{*{7}{|c}|}
\multicolumn{7}{c}{$R_{min} (R/R_{200})$ at different Mass thresholds}\\
\hline
& \multicolumn{2}{c|}{\ahf}  & \multicolumn{2}{c|}{\rockstar} &  \multicolumn{2}{c|}{\subfind} \\
\hline
Level & $10^8 \Mo$ & $10^9 \Mo$ & $10^8 \Mo$ & $10^9 \Mo$ & $10^8 \Mo$& $10^9 \Mo$\\
\hline
1 & 0.04432  & 0.33402 & 0.15591 & 0.33389 & 0.08659 & 0.33395\\
2& 0.07197 & 0.34480 & 0.07190 & 0.37981 & 0.07193 & 0.34472\\
3 & 0.03406 & 0.32940 & 0.00171 & 0.00171 & 0.07818 & 0.32933\\
4 & 0.06800 & 0.37821 & 0.06789 & 0.37827 & 0.11624 & 0.37815\\
5 & 0.13018 & 0.30310 & 0.13065 & 0.30300 & 0.13067 & 0.39331\\
\hline
\end{tabular}
\caption{Values of $R_{min}$ for subhalo mass thresholds of $10^8 \Mo$
and $10^9\Mo$ as a function of resolution level for \ahf, \rockstar\
and \subfind. All the measurements are for the Aq-A halo.}
\label{tab:rmin_vs_lev}
\end{center}
\end{table*}

The same trend appears for all the resolution levels, as seen in Table
\ \ref{tab:rmin_vs_lev}. Here we list the different values of
$R_{min}$ for two different subhalo mass thresholds ($10^8 \Mo$ and
$10^9 \Mo$) at all $5$ resolution levels. These measurements are shown
for \ahf, \rockstar\ and \subfind, the only finders that reach the
highest resolution level. First of all, we can see that the agreement
between finders for the heaviest subhaloes is very good, with the
exception of \rockstar\ at level $3$, where an exceptional subhalo is
found very close to the centre. If we exclude these subhaloes, the \rockstar\ agrees
with the others. However, for the low mass subhaloes the agreement is
not so good. This is not surprising because these small substructures
can move dramatically within the halo when the extra small scale power
is added to the initial power spectrum as the resolution is
increased. This issue particularly affects the central regions of the
halo which are highly non-linear. However, at a fixed resolution level
the finders are trying to extract the same objects. For small masses
we must be careful when we measure the distribution of subhaloes in
the innermost regions as these structures are easy to miss. For large
masses, $R_{min}$ is not only common to all the resolutions but also
in all the finders.

The location of subhaloes within a larger halo and the distribution of
these subhaloes with mass is a consequence of the interplay of the
merging history of the halo, tidal stripping and dynamical
friction. This has important implications for SHAM and other subhalo models. 
Firstly, it is necessary to assess the systematic uncertainties of one's subhalo 
finder as a function of resolution and radius (e.g., such that subhaloes in 
central regions aren't preferentially lost).  Secondly, if one is confident with 
one's subhalo finder, it is necessary to account somehow for the 
subhaloes that have been lost and determine whether `orphan' satellite 
galaxies have survived (e.g., \citealt{Hopkins2010}). 

\subsection{Mass Fractions}\label{sec:mass_frac}

\begin{figure}
\begin{centering}
\includegraphics[width=84mm]{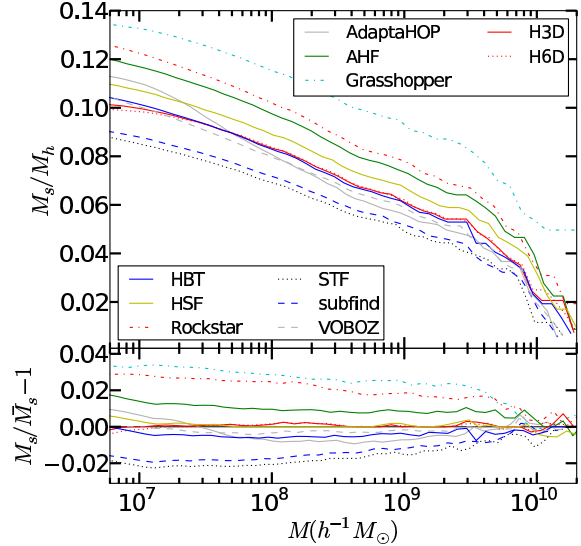}
\caption[Mass Fraction of subhaloes as a function of mass]{Fractional
mass of the host halo in subhaloes for the different finders as a
function of the subhalo mass threshold in the Aq-A halo at resolution
level 4. We restrict this analysis to within $R_{200}$.}
\label{fig:mass_frac_mass} \par\end{centering}
\end{figure}

In Fig.\ \ref{fig:mass_frac_mass} we show the fractional mass of the
host halo that is in subhaloes in the Aq-A halo at resolution level 4
for each of the subhalo finders. The values are shown in terms of mass
 threshold.  We can see that each subhalo finder shows a
different mass fraction, and the differences between them are
approximately constant in the range between $10^7 \Mo$ and $10^9
\Mo$, meaning that the differences are largely due to the size of
the biggest subhaloes. We can see that \grasshopper\ associates a lot of
mass with the largest subhalo.
This is also an important result from the perspective of SHAM models, 
as it implies that the dynamical friction time-scales and merger rates %
inferred from different halo-finding algorithms can vary significantly. %

The radial distribution of subhalo mass has already been examined by
\citet{Onions2012} who show the cumulative mass fraction of subhaloes
as a function of their radial distance from the halo centre. They
found good agreement between the finders except for an excess of
subhaloes for \adaptahop\ .  This work and Fig.~\ref{fig:mass_frac_mass} shows
that most of the finders are apparently consistent in recovering the
masses and radial distances of the subhaloes. This is studied in more
detail in \S \ref{sec:corr_func}.

\subsection{Correlation Functions}\label{sec:corr_func}

\begin{figure*}
\begin{centering}
\includegraphics[width=148mm]{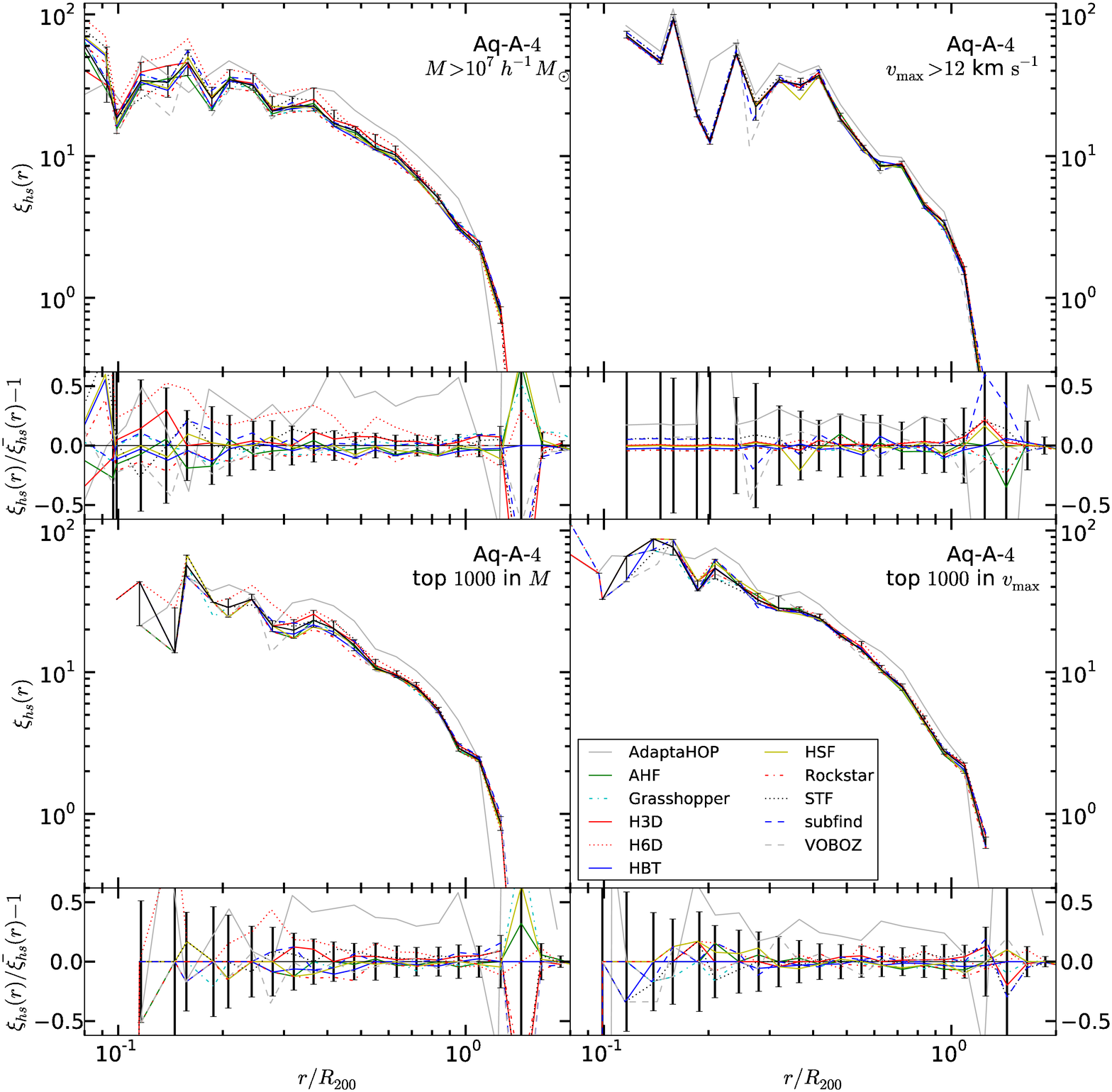}
\caption[Finder comparison of cross CF with mass threshold]{Comparison
between the different finders of the cross CF of subhaloes and the
centre of the halo in Aq-A-4 for subhaloes with $M > 
10^{7} \Mo$ (top left), $v_{\rm max} > 12\ \kms$ (top right
panel), and for top $1000$ subhaloes in mass (bottom left panel) and 
top $1000$ subhaloes in $v_{\rm max}$ (bottom right panel). In the upper part of these 
panels the black line and the error bars correspond to
the median and the $1\sigma$ percentiles respectively. In the
subpanels we show the difference compared to this median. The error 
bars shown in the subpanels represent the poisson shot-noise 
of the AHF finder (all the finders are equivalent) centred in the median. For this
halo $R_{200} = 245.70 \kpc$.} \label{fig:l4_cross_mass}
\par\end{centering}
\end{figure*}

In Fig.\ \ref{fig:l4_cross_mass} we show a comparison of the 
cross CF of subhaloes using different sample cuts. In all the 
subpanels we show the normalized differences of the finders
 with respect to the median, and we show the poisson shot-noise 
 of AHF finder. Since this errors is very similar for all the finders 
 we can assume that this error is a good representation of the shot-noise scatter of these comparisons. 
In the top left panel we show a comparison
of the cross CF of subhaloes with $M_{s} > 10^{7} \Mo$
for the different finders, at resolution level 4. There is a good
level of agreement between the different finders, apart from an excess
in \adaptahop and \hotsixd. These results are consistent with those of \S
\ref{sec:dens_prof} and \S \ref{sec:mass_frac}. On the other hand, we
see from the top right panel of Fig.\ \ref{fig:l4_cross_mass} that the
agreement is even better if we use a $v_{\rm max}$ threshold instead of a
mass threshold. This is because $v_{\rm max}$ is less dependent on the
finder than mass, as shown in Fig.\ 3 and 6 in
\citealt{Onions2012}. This can be explained by the fact that for mass selected subhaloes 
the agreement between the finders depends strongly on how each finder
defines the edge of the subhalo. However the peak of the rotation
curve, $v_{\rm max}$, is defined by the central part of the halo
\citep{Muldrew2011}, so the differences between the finders in
$v_{\rm max}$ are not so strong. In order to make a fair comparison between 
mass and $v_{\rm max}$ cuts we show in bottom panels the finder 
comparison by selecting the top $1000$ subhaloes in mass (left) 
and $v_{\rm max}$ (right). We can see that the agreement between 
finders is stronger when we use the $v_{\rm max}$ cut, although 
in both cases there is a clear excess of \adaptahop. As the results 
using density cuts are similar and present the same conclusions 
than using mass or $v_{\rm max}$ thresholds,  we focus on the mass dependence of
clustering in the paper and we include a study of the $v_{\rm max}$
dependence in the Appendix \ref{sec:vmax_dependence}.

Fig.\ \ref{fig:l4_corr_rvir} shows the 2PCF of subhaloes within
$R_{200}$ and with $M > 10^7 \Mo$ for all the different
subhalo finders in Aq-A in resolution level 4. We can see an 
excess of subhaloes in \hotsixd\ and the fact that the shot-noise 
errors of the 2PCF are smaller.

\subsection{Resolution dependence}\label{sec:resolution_dependence}

In order to see the convergence of the finders with the improving
resolution, we used \subfind, \rockstar\ and \ahf, since they are the
only finders to complete the analysis of all the resolution levels. In
Fig.\ \ref{fig:2pcf_vs_50part_lev} we see how $\xi_{ss}(r)$ depends on
the resolution for these $3$ subhalo finders. In order to avoid
resolution effects due to small haloes at each level, we exclude all
the subhaloes with less than $50$ particles (so each level is resolved
down to a different mass threshold according to Table
\ref{tab:haloes}). The results for level 4 have already been presented
for all finders above. First of all, we can see that levels $4$ and
$5$ present distortions at the smallest scales with respect to the
rest of the levels. So, a high resolution allows us to find subhaloes
with smaller separations between them that we cannot detect at lower
resolution. This is also an indication of the presence of subhaloes
within subhaloes. Moreover, due to the lower subhalo number the shapes
of levels $4$ and $5$ are also more irregular than those of the
highest levels. This can give an idea of the scatter of $\xi_{ss}$ at
these scales due to the resolution of the simulation. As in the
highest resolutions we are including smaller subhaloes, this
comparison is also an indication that the smallest subhaloes smooth
the shape of $\xi_{ss}$.

\begin{figure}
\begin{centering}
\includegraphics[width=84mm]{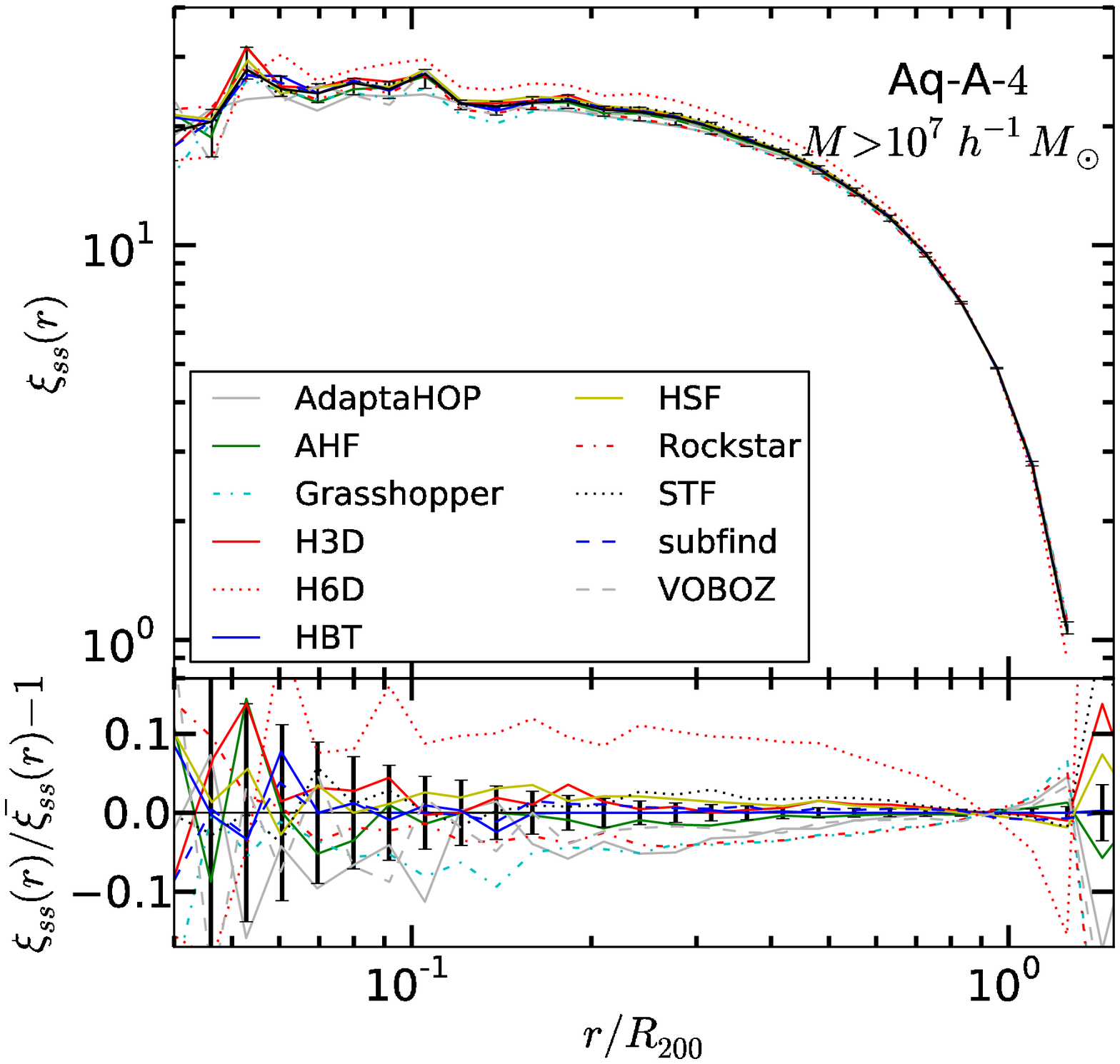}
\caption[Finder comparison of 2PCF with mass threshold
in virialized halo]{Comparison between the different finders for the
2PCF of subhaloes with $M > 10^7 \Mo$ in the Aq-A halo at
resolution level 4. Only subhaloes inside $R_{200}$ are considered. The error bars are obtained as in Fig.\ \ref{fig:l4_cross_mass}. }
\label{fig:l4_corr_rvir} \par\end{centering}
\end{figure}

Surprisingly, finders appear in better agreement at intermediate resolutions.
 The discrepancies from level $5$ are due to the
poor resolution few subhaloes are detected. In
levels $2$ and $3$ the scatter is very low ($\lesssim 5\%$). Given the agreement at this
level we can say that the finders \ahf, \rockstar\ and \subfind\
present very small differences between them when $\xi_{ss}$ includes
all the subhaloes (with more than $50$ particles). This is an
important conclusion, since it means that the definition of subhalo
would not affect the measurements of small scale clustering more than
a few percent, and then measurements of high precision could be
considered reliable. However, in the highest level of resolution the
scatter becomes larger again, up to $10$ percent. This discrepancy
means that finders, at this level of resolution, have different
capabilities of finding small substructure. \rockstar\ finds more
subhaloes with the smallest masses, while \subfind\ tends to be more
conservative and finds less subhaloes. The difference in the
clustering seen in the last panel of Fig.\
\ref{fig:2pcf_vs_50part_lev} can be an indication that these subhaloes
found by \rockstar\ are precisely the most clustered ones. These
differences are only due to the differences in the algorithms of the
finders. In general \subfind\ is one of the most conservative finders,
in the sense that less particles tend to be assigned to the
subhaloes. Then, for a given mass threshold \subfind\ presents fewer
subhaloes. In the centre of the halo the density is higher and the
disruption of the subhaloes is stronger. This can make it difficult to
find small subhaloes unless a strong dynamical analysis is made. On
the other hand, \rockstar\ is designed to produce accurate dynamical
analyses for the structures. This allows the detection of subhaloes which are being
disrupted more easily, and also allows two different subhaloes which
are crossing but not merging to be distinguished. We must also mention
 that some of these extra subhaloes in the centre can be artifacts.  In the end,
\rockstar\ will find more subhaloes in the most clustered regions, and
\subfind\ is designed to be more conservative than the others when
claiming a subhalo detection. From Fig. \ref{fig:2pcf_vs_50part_lev}
we can see that these differences appear in the highest resolution. It
is important to mention that the galaxy distribution within a large
halo is affected by the merging history of the haloes and subhaloes
that make it, and because of this \rockstar\ subhaloes may better
reflect galaxy clustering at these small scales.

We analyse how $\xi_{ss}(r)$ changes with resolution level in Fig.\
\ref{fig:2pcf_lev_mass}, where we see the resolution dependence of
\rockstar\ with different mass thresholds (for $v_{\rm max}$ thresholds
see Fig.\ \ref{fig:2pcf_lev}). The top panel shows $\xi_{ss}(r)$ in
resolution levels 4 to 1 for subhaloes with $M > 10^7 \Mo$. We have
excluded level $5$ since this mass is below $20$ particles at this
level. Although we only show results for \rockstar, the other finders
present similar results. The bottom panel shows the same plot for a
mass threshold of $M > 10^8 \Mo$. First of all, notice that for bottom
panel with the highest mass threshold we cannot see a strong
dependence of $\xi_{ss}(r)$ on resolution. However in the top panel,
with a lower mass threshold, we can see a clear dependence of
clustering on the resolution, showing in the highest level a higher
$\xi_{ss}(r)$, so the results in Aq-A-1 halo have not converged yet. In general, \rockstar\ shows less convergence than the other finders. The resolution dependence in \rockstar\ is stronger and affects larger subhaloes than for the other finders (\ahf\ and \subfind). For these other finders we need to go to smaller subhaloes to see the same effect. 
As at each level we use the same 
threshold, the values of the clustering of these subhaloes are only
dependent on the resolution. In other words, when the resolution is
improved the subhalo clustering changes and an extra term appears in
$\xi_{ss}(r)$. The extra term must come from the smallest subhaloes,
since the largest ones do not show this term. This could be due to the
appearance of small subhaloes included in big structures at the
highest resolution level, which would suppose an indication of a
1-subhalo term in the 2PCF. 

\begin{figure*}
\begin{center}
\includegraphics[width=148mm]{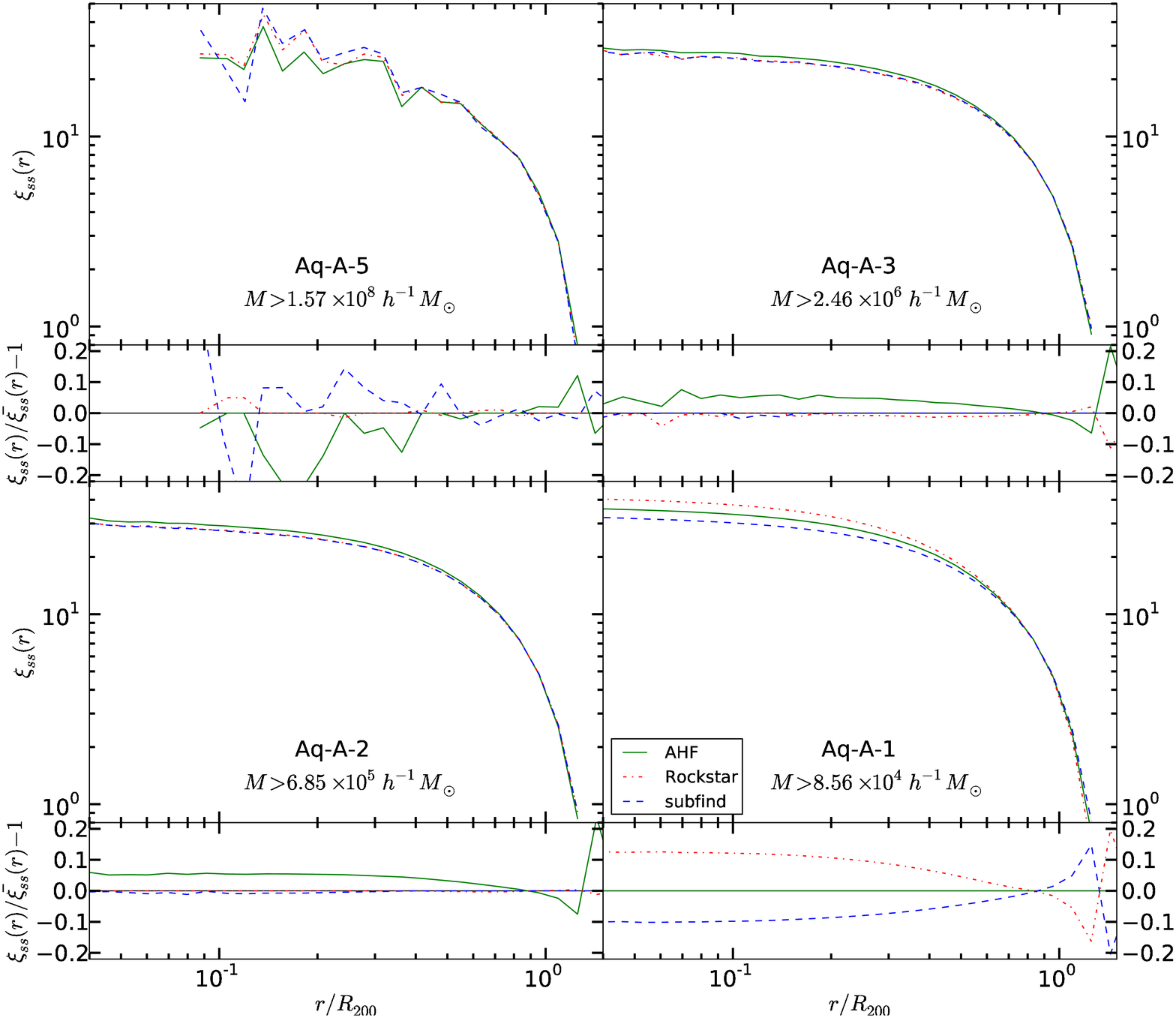}
\end{center}
\caption[2PCF of finders in all levels and with more than 50 part.]
{The subhalo 2PCF of \subfind, \rockstar\ and \ahf\ in Aq-A at resolution levels 5,3,2 \& 1 for subhaloes with more than 50 particles.}
\label{fig:2pcf_vs_50part_lev}
\end{figure*}

This effect is important for the smallest
subhaloes and it can have implications for galaxy clustering. Although 
these subhaloes are small, they can originate from larger subhaloes that 
have lost much of their mass since they were accreted by the host halo. 
This can be important for some halo abundance matching and other halo 
models, because the detection of these subhaloes and their present and 
past properties can complicate the inferred presence and distribution of 
satellite galaxies in the inner regions of haloes.  The differences may 
be important when comparing halo models of galaxy clustering to observed 
clustering at small scales (e.g., \citealt{Wetzel2009}; \citealt{Watson2012}). 

As a consequence of this, the dependence on mass must change with
resolution, since the smallest subhaloes change their clustering
faster than the largest ones. In order to see this explicitly, in Fig.\ \ref{fig:2pcf_mass} we show
the mass dependence of \ahf\ in mass bins for the highest resolution
level 1. We use bins instead of thresholds to see more clearly the
mass dependence of clustering. The results for \subfind\ and
\rockstar\ are similar, although not shown. For the other resolution
levels the dependence on mass is very weak or non-existent. At level
$1$ the smallest subhaloes are the ones with the highest
$\xi_{ss}(r)$. This is due to the fact that, from Fig.\
\ref{fig:2pcf_lev_mass}, the smallest subhaloes increase their
clustering with resolution faster than the most massive ones do. The
effects of Fig.\ \ref{fig:2pcf_lev_mass} and \ref{fig:2pcf_mass} are
stronger if $v_{\rm max}$ dependence is studied instead of mass (see
Figs.\ \ref{fig:2pcf_lev} and \ref{fig:2pcf_vmax}). At some point, the
$\xi_{ss}(r)$ of the smallest subhaloes reaches the $\xi_{ss}(r)$ of
the largest ones, and after that the mass (or $v_{\rm max}$) dependence is
inverted. As this effect is due to the resolution of the simulation,
we can say that at least for resolutions below level $1$ $\xi_{ss}(r)$
is not sensitive to the relation between mass (or $v_{\rm max}$) and
clustering. This is due to the fact that low resolutions are not able
to detect small subhaloes because they simply don't contain enough
particles. This result is important since there are no large
scale simulations nowadays with the resolution of level $1$, and for all
these simulations we could be underestimating the clustering of the
smallest subhaloes. If the difference is because of the inclusion of
the substructure of subhaloes, we can say that one of the effects of
including this 1-subhalo term is the inversion of the mass (or
$v_{\rm max}$) dependence on clusteringat these scales.

\begin{figure}
\includegraphics[width=84mm]{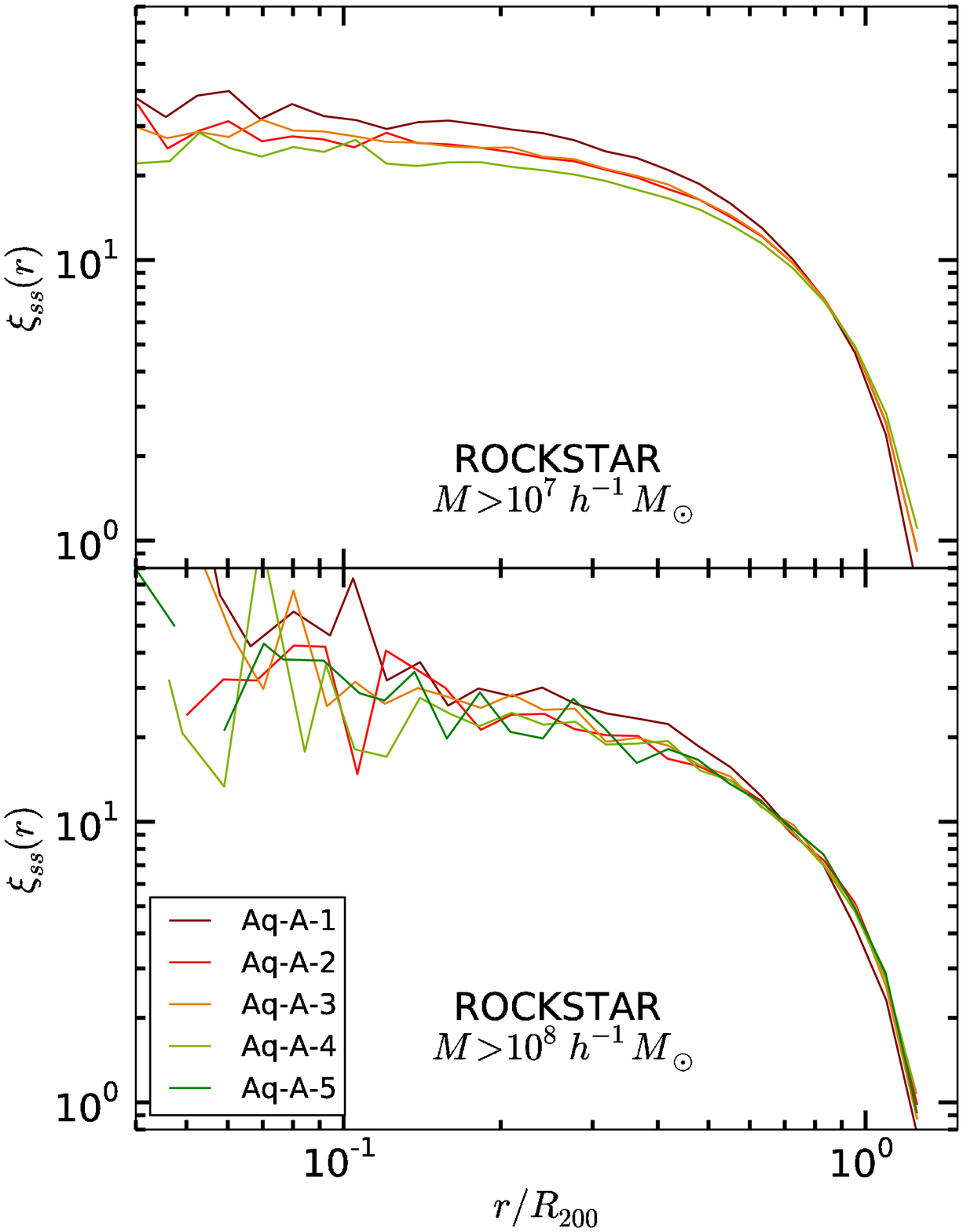}
\caption[2PCF of \rockstar\ vs level with different mass thresholds.]
{The subhalo 2PCF of \rockstar\ in Aq-A at 4 different resolution
levels for two mass thresholds. In the top panel, the sample
corresponds to subhaloes with $M> 10^7 \Mo$. The bottom panel shows
subhaloes with $M > 10^8 \Mo$.}
\label{fig:2pcf_lev_mass}
\end{figure}

\begin{figure}
\includegraphics[width=84mm]{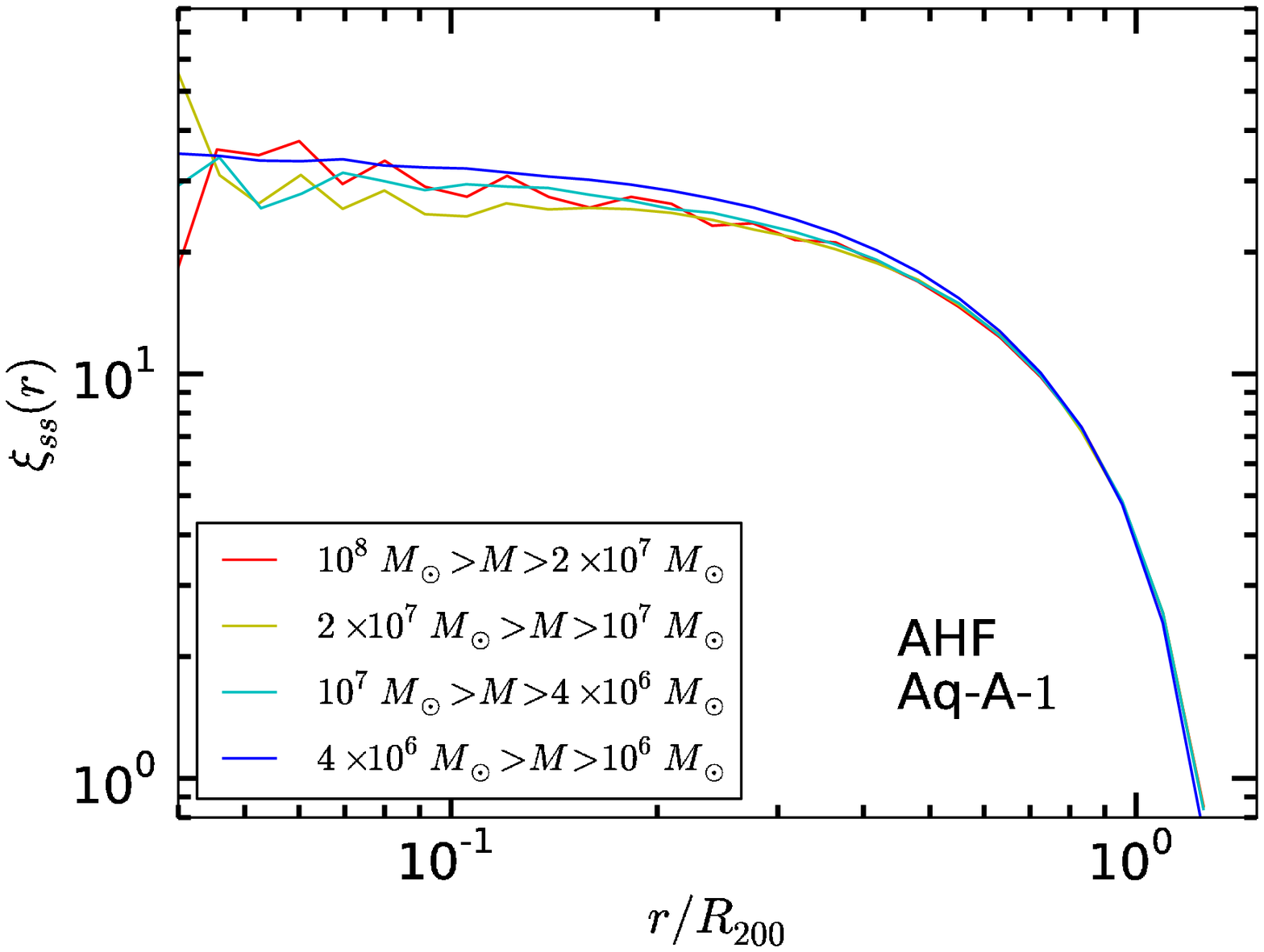}
\caption[2PCF of \ahf\ as a function of mass for level $1$.]
{2PCF of subhaloes from \ahf\ for different mass thresholds as indicated in the panel for Aq-A resolution level 1.}
\label{fig:2pcf_mass}
\end{figure}

\section{Halo Model}

To describe the 2PCF and cross CF analytically, we follow the extended
halo-model    formalism    developed    by    \citet{Sheth2003}    and
\citet{Giocoli2010b}.  Since  real space convolutions  are represented
by multiplications in  Fourier space, we first write  the equations of
the power spectra and then convert them to real space:
\begin{equation}
\xi(r) = \dfrac{1}{2 \pi^2} \int k^3 P(k) \dfrac{\sin(k r)}{k r} 
\dfrac{\mathrm{d} k}{k}\,.
\end{equation}

As  in  the  halo  model  formalism  for  the  matter  power  spectrum
reconstruction
\citep{Scherrer1991,Seljak2000,Scoccimarro2001,Cooray2002},        the
subhalo-subhalo  power  spectrum can  be  split  into a  Poisson  term
$P_{ss,1H}(k)$  that describes  the  contribution  from subhalo  pairs
within an  individual halo (the  `one-halo term'), plus  a large-scale
term $P_{ss,2H}(k)$ that describes  the contribution from subhaloes in
separate haloes (the `two-halo term'):
\begin{equation}
P_{ss}(k) = P_{ss,1H}(k) + P_{ss,2H}(k)\,.
\label{eqpkss}
\end{equation}

Both  terms   require  knowledge   of  the  subhalo   spatial  density
distribution, and the halo and subhalo mass functions.  In particular,
the second term requires a model for the power spectrum of haloes with
different mass $P_{hh}(k|M_1,M_2)$ that with good approximation can be
expressed as a function of the  halo bias $b(M)$ and the linear matter
power spectrum $P_{\rm lin}(k)$:
\begin{equation}
P_{hh}(k|M_1,M_2) \approx b(M_1) b(M_2) P_{\rm lin}(k)\,.
\end{equation} 

To model  the spatial density  distribution of subhaloes  $n_s$ around
the center of a halo  with mass $M$ and concentration $c=r_s/R_{200}$,
needed in  the reconstruction  of both the  one-halo and  the two-halo
term  in  equation~(\ref{eqpkss}),  we adopt  the  analytical  fitting
function by \citet{Gao2004}, motivated by  an analysis of results from
numerical simulations:
\begin{equation}
f(<r|c) = \dfrac{n_s(<x|c)}{N_{tot}(M|c)} = 
\dfrac{(1+ \alpha c) x^{\beta} }{1+ \alpha c x^2}\,,
\end{equation}
where  $x$ is  the  distance from  the center  in  unit of  $R_{200}$,
$\alpha=0.244$, $\beta=2.75$  and $N_{tot}(M|c)$ represents  the total
number of subhaloes within $R_{200}$, that we assume to depend both on
host           halo          mass           and          concentration
\citep{Delucia2004,Vandenbosch2005,Giocoli2008,Giocoli2010}.   It  has
been observed that,  at a given redshift, more massive  haloes host on
average more  substructures than less  massive ones, because  of their
lower formation redshift;  in addition, at a fixed  mass and redshift,
more concentrated haloes host  fewer structures than less concentrated
ones \citep{Gao2008}.  To compute  the normalized Fourier transform of
the  subhalo  distribution  in  a  spherically  symmetric  system,  we
numerically solve the equation:
\begin{equation}
u_s(k|c)  =  \int_0^{R_{200}}  4  \pi  r^2
\dfrac{\sin k r}{k r} f_s(r|c) \mathrm{d} r,
\end{equation}
where  $f_s(r|c)$  represents   the  normalized  differential  subhalo
density  distribution  around the  host  halo  center, i.e.  with  the
condition that $f_s(<R_{200})=1$.

Now  we can  write  the  one-halo and the two-halo term  of  the subhalo-subhalo  power
spectrum as follows:
\begin{equation}
\begin{aligned}
P_{ss,1H}(k) & = \int_M \dfrac{n(M)}{\bar{N}_{tot}^2} \times \\  
& \int_c N^2_{tot}(M|c) \,u^2_s[k|c(M)] \, p(c|M) \mathrm{d} c \, \mathrm{d} M,
\end{aligned}
 \label{eq:ss1H}
\end{equation}

\begin{equation}
\begin{aligned}
 \label{eq:ss2H} P_{ss,2H}(k) & = P_{lin}(k)  \left[ \int_M  
\dfrac{n(M) b(M)}{\bar{N}_{tot}} \right. \times \\ & \left. \int_c
N_{tot}(M|c) \, u_s[k|c(M)] p(c|M) \mathrm{d} c \, \mathrm{d} M \right]^2.
\end{aligned}
\end{equation}

where $n(M)$ is  the halo mass function,  $\bar{N}_{tot}$ the comoving
mean  number  density of  satellites  in  the universe,  $p(c|M)$  the
log-normal  scatter  in  concentration  at fixed  halo  mass,  and  we
explicitly express  the mass dependence of  concentration in $u_s(k)$.

For the halo-subhalo \textit{cross} power spectrum we have, respectively:
\begin{equation}
\begin{aligned}
P_{hs,1H}(k) & = \int_M \dfrac{n(M)}{\bar{N}_{tot} \bar{N}_{h,tot}} \times  \\  
& \int_c N_{tot}(M|c) \,u_s[k|c(M)] \, p(c|M) \mathrm{d} c \, \mathrm{d} M,
 \label{eq:hs1H}
\end{aligned}
\end{equation}
and 
\begin{equation}
\begin{aligned}
P_{hs,2H}(k) &= P_{lin}(k) \int_{M_1} \dfrac{n(M_1) b(M_1)}{\bar{N}_{tot}} \times \\ \int_c
&N_{tot}(M_1|c) \, u_s[k|c(M_1)] p(c|M_1) \mathrm{d} c \, \mathrm{d} M_1 \times \\  
& \int_{M_2} \dfrac{n(M_2) b(M_2)}{\bar{N}_{h,tot}} \mathrm{d} M_2,
 \label{eq:hs2H}
\end{aligned}
\end{equation}
where $\bar{N}_{h,tot}$ represents the comoving mean number density of
haloes   in    the   Universe.    Note   that    (\ref{eq:ss1H})   and
(\ref{eq:hs1H}), respectively, can be thought of as contributions from
satellite-satellite   and    center-satellite   terms   in    a   halo
\citep{Sheth2005,Skibba2006}.

Since we  are measuring the  2PCF and  CF of the  subhalo distribution
within  $R_{200}$  in a  single  halo,  the  important terms  will  be
\textit{only}  the  Poisson  ones  $P_{ss,1H}$  and  $P_{sh,1H}$;  the
large-scale terms (\ref{eq:ss2H}) and (\ref{eq:hs2H}) appear only when
measurements in a simulation can be extended out to larger separations
($r \gg R_{200}$).

\begin{figure}                                                                  
\begin{centering}                                                               
\includegraphics[width=\hsize]{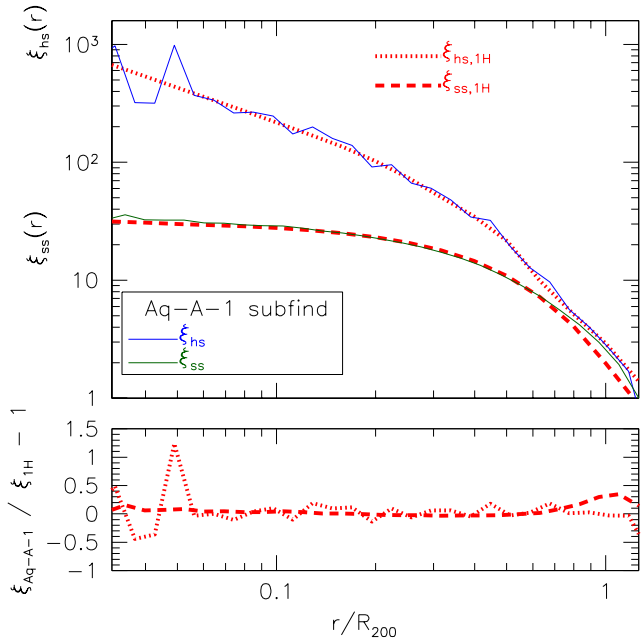}       
\caption[]{Comparison   of    the   two-point    autocorrelation   and
  cross-correlation functions of the extended halo-model formalism and
  the measurements  of the  Aq-A-1 run  with the  subhaloes identified
  with \subfind\ (similar  to Fig.~1).  The solid green  and blue lines
  show the 2PCF  and cross CF of the simulation,  while the red dashed
  and dotted curves show the halo-model predictions, respectively. In the bottom frame
  we show the residuals of the measurements with respect to the halo model prediction.}
\label{fig:hmpred}                                 
\par\end{centering}                                                             
\end{figure}

In Fig.~\ref{fig:hmpred} we compare this halo-model description of the
one-halo terms of $\xi_{ss}(r)$ and $\xi_{hs}(r)$ to the measured 2PCF
and cross CF of  the Aq-A-1 run.  For this purpose,  we have chosen to
compare our analytical  predictions to the \subfind\  catalogue, as the
spatial subhalo density distribution model of \citet{Gao2004} has been
tuned to a  set of simulations in which subhaloes  are identified with
the same algorithm.  For the halo  and subhalo mass functions, we have
adopted   those   from  \citet{Sheth1999}   and   \citet{Giocoli2010},
respectively.  We  have used  the same halo  parameters of  the A-Aq-1
halo --  the integrals on the  mass function and on  the concentration
distribution are  restricted around  the halo  parameters of  the halo
(see  Table~\ref{tab:haloes})   --  and  have  adopted   an  arbitrary
normalization,  as was  done  for the  measurements  in the  numerical
simulation.

There is  clearly very good  agreement between the halo  model prediction and
the  simulation,   for  both  the  subhalo-subhalo   and  halo-subhalo
clustering signals, over  a wide range of scales.   This lends support
for the model, which provides a good description for the abundance and
distribution of  subhaloes around a  host halo, as calibrated by \citet{Gao2004} 
for cluster-size haloes and extended in this work to the Aquarius simulation.  It  also demonstrates
that this simulation  does not contain an atypical halo,  in the sense
that  its  substructures appear  to  be  consistent with  the  average
clustering  properties   of  multiple  haloes  in   other  simulations
\citep{Gao2004,Giocoli2010}.

\section{Discussion and conclusions}\label{sec:conclusions}

Using a diverse set of subhalo finders, we studied the radial distribution of subhaloes inside Milky Way-like dark matter haloes as simulated within the framework of the Aquarius project. Our interest was focused on the number density profile and two-point correlation functions (2PCF), respectively, investigating any possible variations coming from the utilization of distinct finders as well as the convergence of the results across the different resolution levels of the simulation itself. This work forms part of our on-going "Subhalo Finder Comparison Project" described in greater detail elsewhere \citep[e.g.][]{Knebe2013}. And following the spirit of the previous comparisons, each code was only allowed to return a list of particle IDs from which a common post-processing pipeline calculated all relevant subhalo properties, including the position. However, this pipeline does not per-se feature an unbinding procedure which was still left to the actual finder in this study. \\

Our principal conclusions can be summarized as follows:

(i) The number density profile and radial distribution of subhaloes in the Aquarius haloes is different to the underlying dark matter density profile described by the functional form proposed by NFW \citep{Navarro1996}. This is an important result, since in many studies and observations one assumes a NFW profile, also for the subhalo distribution. Even more, a deficit of subhaloes in the central regions of the host actually led to the introduction of so-called 'orphan galaxies' in order to bring observations into agreement with simulations \citep{Springel2001,Gao2004b,Guo2010,Frenk2012}: it can and does happen that a dark matter subhalo dissolves due to tidal forces (and lack of numerical resolution) while orbiting in its host halo \citep[e.g.][for a study of these disrupted subhaloes]{Gill2004b}. However, a galaxy having formed prior to this disruption and residing in it should survive longer than this subhalo. Therefore, it became standard practice to keep the galaxy alive even though its subhalo has disappeared, calling it 'orphan galaxy'.

(ii) The number density profile of subhaloes depends on the mass of the subhalo. For each subhalo mass, we have found a minimum distance from the halo centre that increases with subhalo mass (cf. Fig.\ \ref{fig:r_min_vs_mass}). This is due to the fact that massive subhaloes are rarer, and then it is difficult to find one of them close to the centre. But we also caution the reader that this result is weakened by the fact that practically every halo finder reduces the subhalo mass when placed closer to its host centre (cf. upper panel of Fig.8 in \cite{Knebe2011} and Fig. 4 in \cite{Muldrew2011}).

(iii) The subhalo finders differ considerably on the fraction of mass in subhaloes with the deviations primarily driven by the most massive subhaloes (cf. Fig.~\ref{fig:mass_frac_mass}). We also confirmed (though not explicitly shown here) that most of the mass in subhaloes is localized outside $0.4 R_{200}$, consistent with the previous result that the most massive subhaloes are found far from the halo centre. 

(iv) All codes show a remarkable agreement on the cross CF and 2PCF inside the radius $R_{200}$.  With the exception of \adaptahop\ finder, in most of the cases the agreement between finders is consistent with Poisson shot-noise errors. For the 2PCF using mass bins, we find $10\%$ agreement between finders for $r > 0.1 R_{200}$, although the shot-noise error decreases faster with $r$. This reassures us that correlation measurements inside (the virial part of) haloes are not influenced by the choice of the finder at the $10\%$ level of accuracy.

(v) However, we did find that for both the lowest and highest resolution levels there are differences amongst the finders. The former can be attributed to poor resolution, whereas the latter clearly reveals differences in the codes: while the contrast for subhaloes in the very central regions increases some finders still struggle to detect those objects flying past the innermost centre of the host. Further, the highest resolution level shows clear signs of sub-subhaloes yet another possible challenge for halo finders.

(vi) The 2PCF of small subhaloes depends strongly on resolution, with increasing clustering for increasing resolutions. This effect is stronger when a $v_{\rm max}$ threshold is applied and for small subhaloes. For $v_{\rm max} > 7\ \kms$, the 2PCF increases between $10\%$ and $30\%$ (with a shot-noise uncertainty of $5-10\%$) at the highest resolution. This can be an indication of an extra term of the 2PCF detected in the highest resolution, probably the $1$-\textit{sub}halo term of the correlations (i.e. the existence of sub-subhaloes). As this effect is stronger for the smallest subhaloes, the mass dependence of clustering depends on the resolution, too. In particular, we see that at level $1$ there is an anti-correlation between clustering and mass. The importance of this result resides in the fact that, as there are no large scale simulations with this resolution to-date, the clustering of the smallest subhaloes in these simulation can be systematically underestimated. 

(vii) We confirm aforementioned findings when using $v_{\rm max}$ as opposed to mass cuts (cf. Appendix~\ref{sec:vmax_dependence}), albeit a stronger dependence on $v_{\rm max}$. When a $v_{\rm max}$ cut is applied, the difference between finders is smaller than the $5\%$ level for $r > 0.2 R_{200}$, an agreement consistent with the Poisson shot-noise. As $v_{max}$ retains more information about the past of the subhaloes and provides a more suitable measure when comparing to observations, this result will have more importance for the implications in galaxy formation. 

All these results certainly contribute to the understanding of the 
substructure distribution within dark matter haloes and how much their distribution 
depends on the finder algorithm and the resolution of the 
simulation. Moreover, substructure clustering plays an important role for 
galaxy formation models, because satellite galaxies are expected to
follow the subhalo gravitational potentials. Methods such as 
Sub-Halo Abundance Matching (SHAM) often make this assumption, and 
Semi-Analytical Models (SAM) model baryonic processes according to the 
properties of the subhaloes and their merger trees. In these cases, the 
properties of the subhaloes affect inferred properties of the galaxy 
population, and a correct definition/identification of subhaloes is 
crucial. For a given model, using different subhalo finders can produce 
different galaxy distributions  and galaxy-subhalo relations. These 
differences make it difficult to compare galaxy formation models and their 
predictions if their assumptions about subhalo definition and 
identification are treated differently.

Also, HOD models populate galaxies in simulations to infer halo properties from observations, usually a NFW profile of the galaxies in haloes is assumed. But if galaxies follow the subhalo distribution instead of the dark matter field, then the measurement of the subhaloes can also be used in the HOD models to improve the radial distribution of galaxies in haloes. 

Improving the resolution of the simulation is also crucial, since we have 
seen that many subhaloes are lost in the lowest resolution simulations, 
and they have important consequences for the resulting subhalo clustering. 
Most of these lost subhaloes are rather small and live in the densest 
regions of their host, but in these cases they could have been more 
massive and experienced severe tidal stripping, respectively. 
If one uses abundance matching to populate galaxies in simulations when 
the resolution is insufficient, if one is using present subhalo 
mass or $v_\mathrm{max}$ (or also these quantities at the time of 
accretion), one could be missing an important fraction of galaxies in the 
centre of the haloes, and workers in the field try to circumvent this by 
introducing aforementioned orphan galaxies \citep{Springel2001,Gao2004b,Guo2010,Frenk2012}.
It is therefore important to consistently track haloes and subhaloes 
when associating them with galaxy populations. 

As already highlighted previously \citep{Onions2013}, the (non-)removal of unbound particles will leave an impact on subhalo properties, too -- something again confirmed in this work: we have found that \adaptahop\  show important differences to the rest of the finders, since it does not include a (faithful) removal of unbound particles. \adaptahop\ does not eliminate the background particles from the host  -- unbound to the subhalo -- which produces an overestimation of the number of small subhaloes, especially in the central parts of the halo. This effect clearly leaves an imprint in the number density profile and correlation functions.

The implications of our results also extend to the interpretation of ongoing and upcoming galaxy surveys measuring a fair fraction of the observable Universe (just to name a few, BOSS, PAU, WiggleZ, eBOSS, BigBOSS, DESpec, PanSTARRS, DES, HSC, Euclid, WFIRST, etc.). For their interpretation of the 2-point galaxy correlation function (or alternatively the power spectrum) is commonly used and hence needs to be determined to unprecedented accuracy \citep[e.g.][]{Smith2013}. As we have just seen, the 1-halo and in particular the 1-subhalo term is sensitive to the applied halo finder. Further work and 
analysis in high resolution cosmological simulations are needed to better understand this.

But all our results have to be taken with a grain of salt: as one single halo does not represent a homogeneous distribution, we must be careful with the definition and interpretation of the 2PCF. We use a theoretical normalization where we assume an infinite and completely homogeneous random field. The results converge to the random sample normalization when the volume of the random sample is large enough. We need to assume an arbitrary mean density of subhaloes, since we cannot measure the abundance of subhaloes expected for a large simulation because there are no large simulations with the resolution of these Aquarius haloes. This measurement of the 2PCF must be understood as the contribution that the halo would give to the $1$-halo term of the 2PCF of a large and homogeneous simulation, since it reflects the number of pairs of subhaloes found inside the halo, with an arbitrary amplitude due to the unknown mean number density of subhaloes.

\section*{Acknowledgements}

The work in this paper was initiated at the ‘ Subhaloes going Notts’ workshop in Dovedale, which was funded by the Euro- pean Commissions Framework Programme 7, through the Marie Curie Initial Training Network CosmoComp (PITN-GA-2009- 238356). We wish to thank the Virgo Consortium for allowing the use of the Aquarius data set.

The authors contributed in the following ways to this paper: AP, EG, CG, AK, FRP, RAS undertook this project. They performed the analysis presented and wrote the paper. AP is a PhD student supervised by EG. FRP, AK, HL,  JO and SIM organized and ran the workshop at which this study was initiated. They designed the comparison study and planned and organized the data. The other authors provided results and descriptions of their algorithms as well as having an opportunity to proof read and comment on the paper.

The authors wish to thank Ravi Sheth for valuable discussions about modeling subhalo correlation functions.

YA receives financial support from project AYA2010-21887-C04-03 from the former Ministerio de Ciencia e Innovación (MICINN, Spain), as well as the Ramón y Cajal programme (RyC-2011-09461), now managed by the Ministerio de Economía y Competitividad.
CG's research is part  of the project GLENCO, funded under the European  Seventh  Framework  Programme,  Ideas,  Grant  Agreement  n.259349.
AK is supported by the MICINN in Spain through the Ram\'{o}n y Cajal programme as well as the grants AYA 2009-13875-C03-02, AYA2009-12792-C03-03, CSD2009-00064, CAM S2009/ESP-1496 (from the ASTROMADRID network) and the {\it Ministerio de Economía y Competitividad} (MINECO) through grant AYA2012-31101. He further thanks East River Pipe for the gasoline age.
HL acknowledges a fellowship from the European Commission’s Framework Programme 7, through the Marie Curie Initial Training Network CosmoComp (PITN-GA-2009-238356).
A.P. is supported by beca FI from Generalitat de Catalunya. Funding for this project was partially provided by the MICINN, project AYA2009-13936, Consolider-Ingenio CSD2007- 00060, European Commission Marie Curie Initial Training Network CosmoComp (PITN-GA-2009-238356), research project 2009- SGR-1398 from Generalitat de Catalunya. 
RAS is supported by the NSF grant AST-1055081.
SIM acknowledges the support of the STFC Studentship Enhancement Program (STEP).

\bibliography{aamnem99,biblist}

\appendix

\section{$v_{\rm max}$ dependence}
\label{sec:vmax_dependence}

\begin{figure}
\begin{centering}
\includegraphics[width=84mm]{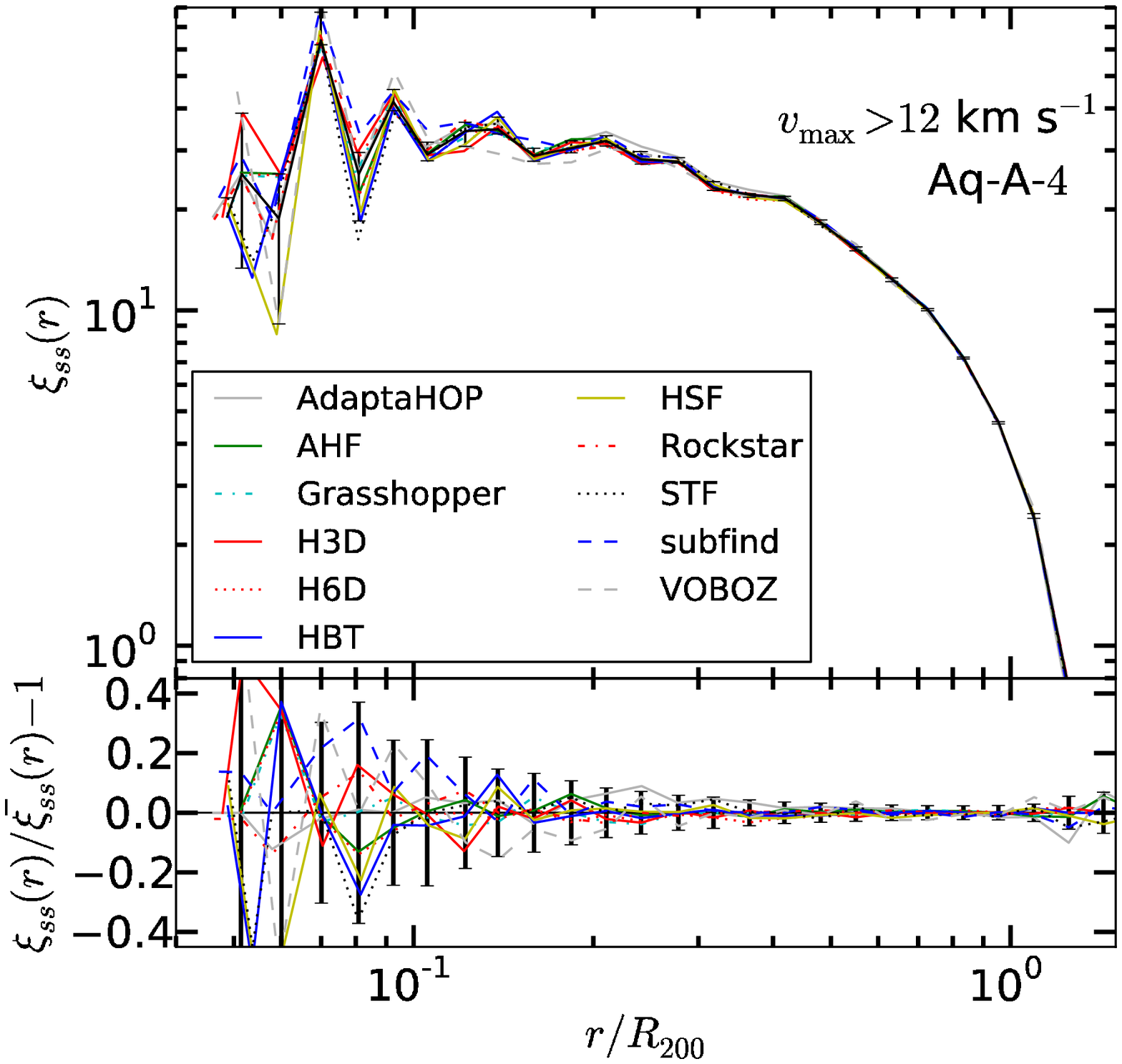}
\caption[Finder comparison of 2PCF with $v_{\rm max}$ threshold]{Comparison between the different finders of the 2PCF of subhaloes with $v_{\rm max} > 12\ \kms$ in the Aq-A halo at level 4 of resolution. The black line and the error bars in the upper part correspond to the median and the $1 \sigma$ percentiles respectively, while in the bottom subplot the error bars represent the Poisson shot-noise of the \ahf\ finder (the other finders are equivalent).} \label{fig:l4_corr_vmax}
\par\end{centering}
\end{figure}

\begin{figure*}
\includegraphics[width=148mm]{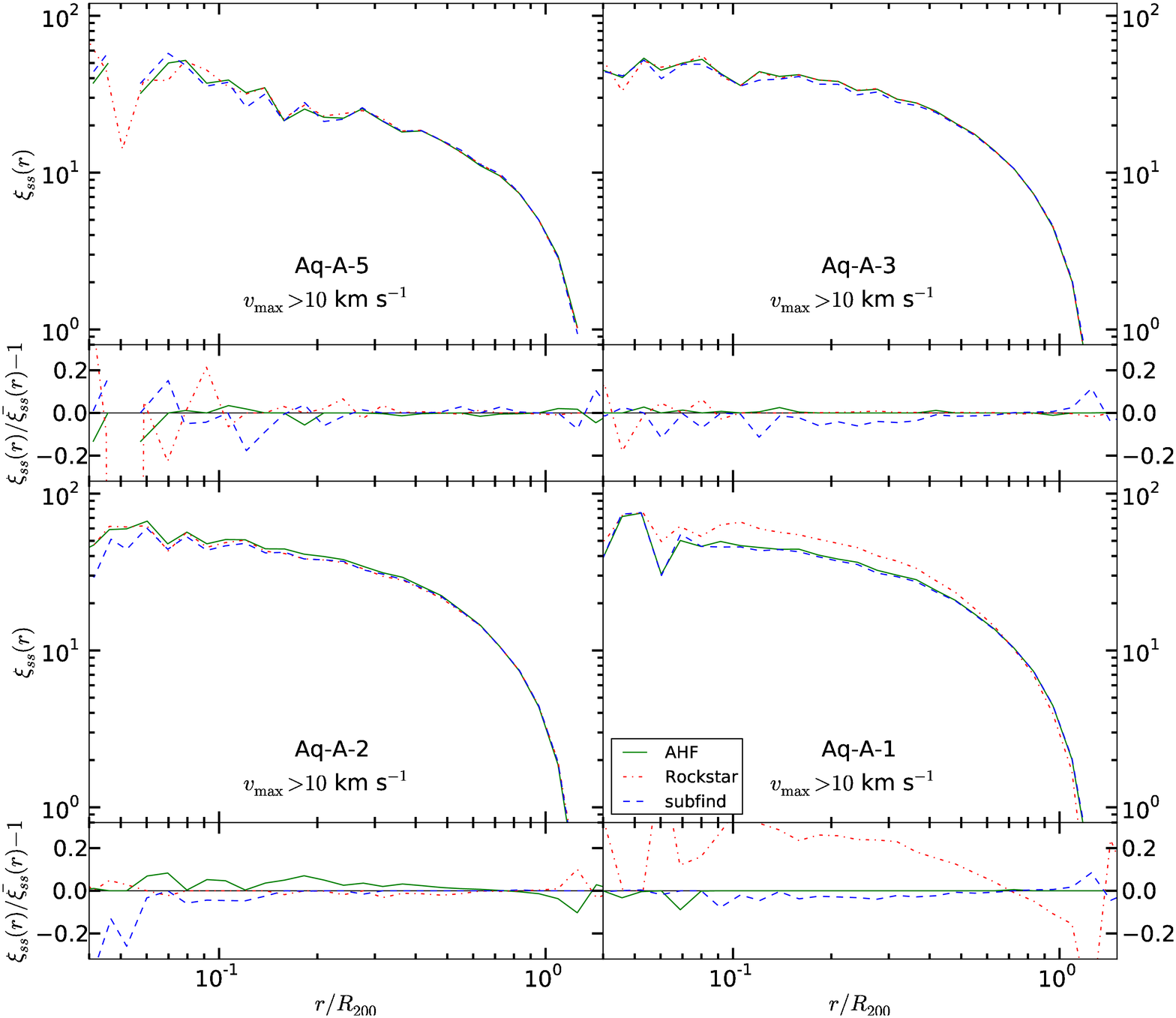}
\caption[2PCF of finders in all the resolution levels and with $v_{\rm max} > 10\ \kms$.]
{The subhalo 2PCF of \subfind, \rockstar\ and \ahf\ in Aq-A at different resolution levels and $v_{\rm max} > 10\ \kms$ threshold.}
\label{fig:2pcf_vs_vmax_lev}
\end{figure*}

\begin{figure}
\includegraphics[width=84mm]{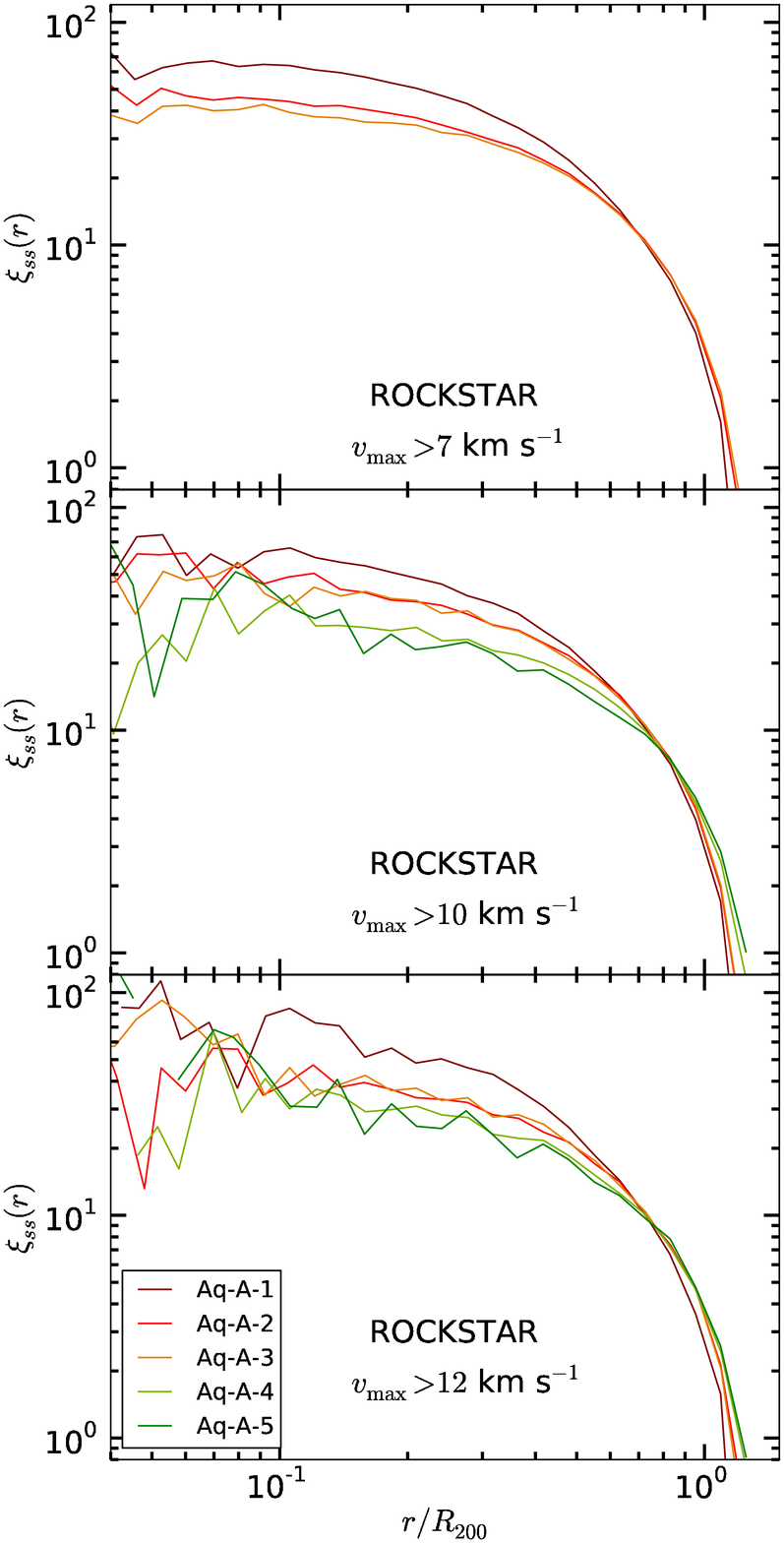}
\caption[2PCF of AHF vs level with $v_{max }> 7\ \kms$, $v_{max }> 10\ \kms$  and $v_{\rm max} > 12\ \kms$.]
{The subhalo 2PCF of \rockstar\ in Aq-A as a function of the resolution levels for $3$ $v_{\rm max}$ thresholds. In the top panel, the sample corresponds to subhaloes with $v_{\rm max} > 7\ \kms$. In the middle the threshold used is $v_{\rm max} > 10\ \kms$. Finally, in the bottom panel a threshold of $v_{\rm max} > 12\ \kms$ has been used.}
\label{fig:2pcf_lev}
\end{figure}

\begin{figure}
\includegraphics[width=84mm]{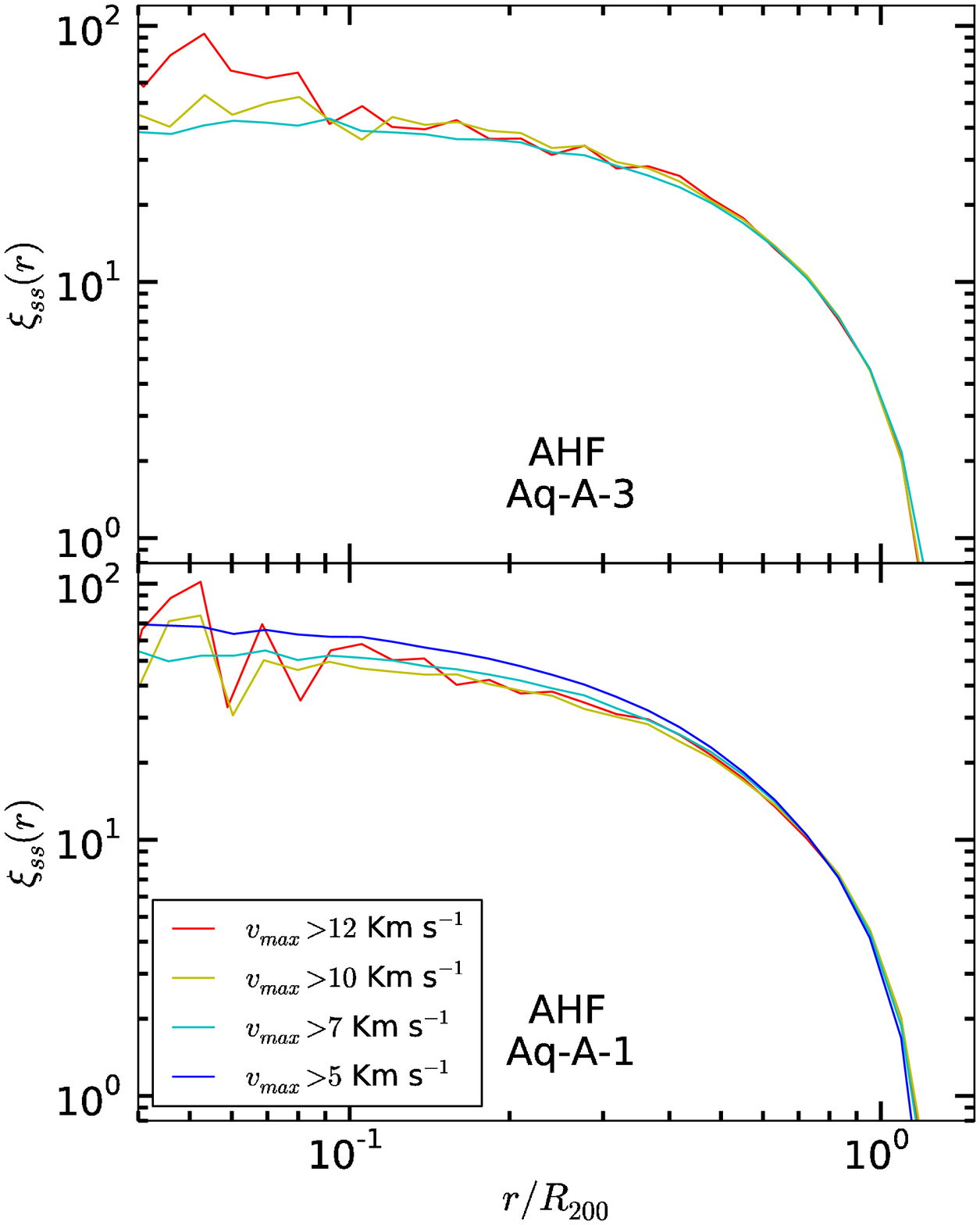}
\caption[2PCF of AHF vs $v_{max }$ at levels $1$, $2$ and $3$ ]
{2PCF of \ahf\ at different $v_{\rm max}$ thresholds, at levels $1$ (bottom) and $3$ (top).}
\label{fig:2pcf_vmax}
\end{figure}

In this Appendix we explore the same study as in \S \ref{sec:results} but using $v_{\rm max}$ instead of mass as the subhalo property studied.

In Fig.\ \ref{fig:l4_corr_vmax} we see $\xi_{ss}(r)$ for subhaloes with $v_{\rm max} > 12\ \kms$. However, the agreement is much better for $v_{\rm max}$ thresholds than for mass threshold. This is constant in all the study. We can see in particular that the differences between finders are consistent with the shot-noise errors due to their statistics.

In Fig. \ref{fig:2pcf_vs_vmax_lev} we compare the finders \ahf\, \rockstar\ and \subfind\ in several resolution levels with the $v_{\rm max}$ threshold $v_{\rm max} > 10\ \kms$. We do not show level $4$ since it is shown in Fig.\ \ref{fig:l4_corr_vmax}. First of all, we can see a strong scatter at levels $4$ and $5$ in scales lower than $0.1 R_{200}$. This, as in the case of mass thresholds, can be understood from the difficulty of finding subhaloes with small separations at these levels of resolution. Then, these resolutions are not sensitive to the 2PCF of subhaloes at these scales. On the other hand, at larger distances the agreement of the finders at these levels as well as at levels $2$ and $3$ is remarkable. However, the differences become stronger at level $2$ and much larger at level $1$ of resolution. In particular, in level $1$ \rockstar\ shows a large difference between the other finders. Although the results are equivalent to Fig.\ \ref{fig:2pcf_vs_50part_lev} where we have used a mass threshold, we find a strong disagreement  between  ROCKSTAR and the other finders when the $v_{\rm max}$ threshold is applied at level $1$. This is an indication that the extra subhaloes found by \rockstar\ are precisely those with more clustering as $v_{\rm max}$ is more sensitive to the clustering of these subhaloes than mass.

As we see, $\xi_{ss}(r)$ changes with resolution also for $v_{\rm max}$ thresholds. We can see these changes more explicitly in Fig.\ \ref{fig:2pcf_lev}, where we show $\xi_{ss}(r)$ for \rockstar\ finder as a function of the resolution using $3$ different $v_{\rm max}$ thresholds ($v_{\rm max} > 7\ \kms$ on top, $v_{\rm max} > 10\ \kms$ in the middle and $v_{\rm max} > 12\ \kms$ in the bottom panel). We only show one finder, but the results are equivalent for the others. For each threshold we only show the levels of resolution that are consistent with this threshold. First of all, we can see that the regularity in the shapes of $\xi_{ss}(r)$ is improved for higher resolutions, meaning that this irregularity is purely due to resolution effects. Secondly, we note that the clustering is higher for higher resolutions in all the thresholds used. We can see this effect clearer using $v_{\rm max}$ thresholds instead of mass thresholds because $v_{\rm max}$ is more strongly related to clustering than mass. As for mass thresholds, \rockstar\ shows a larger effect than the other finders. Then, in Fig.\ \ref{fig:2pcf_lev} we can see that an extra term appears in $\xi_{ss}(r)$ when we improve the resolution. This means, again, that simulations with lower resolutions are not able to appreciate this extra term. Subhaloes with low $v_{\rm max}$ increase faster their clustering with resolution than subhaloes with large $v_{\rm max}$, as we will see in Fig.\ \ref{fig:2pcf_vmax}. This might be an indication of the sub-substructure detected only in the highest resolutions, as discussed in \S \ref{sec:resolution_dependence}. 

Finally, we analyse the $v_{\rm max}$ dependence of subhalo clustering in Fig.\ \ref{fig:2pcf_vmax}. Fig.\ \ref{fig:2pcf_vmax} shows the clustering dependence on $v_{\rm max}$ of AHF finder at levels $1$ (bottom) and $3$ (top). Although the interpretation in the smallest scales is complicated due to the low statistics, the results at larger scales are clear.  At level $1$, we see that the lower the threshold, the higher the clustering. However, at level $3$ the relation is the opposite than in level $1$. This result shows that the smallest subhaloes increase their clustering with resolution faster that the largest ones do. The change on clustering is higher if the subhaloes are smaller, and they change up to the point of inverting the relation between clustering and $v_{\rm max}$ from level $3$ to $1$. As the relation between $v_{\rm max}$ and clustering is stronger than mass, the change in the $v_{\rm max}$ dependence is stronger an easier to see that the change on the mass dependence shown in Fig.\ \ref{fig:2pcf_mass}. Again, the appearance of an extra term on the 2PCF of the smallest subhaloes can be an indication of the detection of the $1$-subhalo term of the 2PCF. The fact that this effect also happens for $v_{\rm max}$ threshold is important, since $v_{\rm max}$ is expected to retain more information about the history and past of the subhaloes than mass. The distribution of subhaloes as a function of $v_{\rm max}$ is important for methods of galaxy formation such as SHAM, where subhalo $v_{\rm max}$ has been shown to be a better tracer of galaxies than subhalo mass \citep{Reddick2013,Hearin2013}. The conclusions made from the mass dependence of this 1-subhalo term about the implications on SHAM are more important when $v_{\rm max}$ is taken into account, not only because $v_{\rm max}$ reflects more galaxy clustering on SHAM galaxies, but also because this effect is even stronger.

\label{lastpage}

\end{document}